\documentclass[useAMS,usenatbib]{mn2e}

\usepackage{latexsym}
\usepackage{amsmath}
\usepackage{amssymb}
\usepackage{graphicx}
\title[On the star-forming ability of Molecular Clouds]{On the star-forming ability of Molecular Clouds}
\author[Anathpindika. S, Burkert. A and Kuiper, R.]{Anathpindika, S $^{1},^{2},^{3}$\thanks{E-mail: sumed$\_$k@yahoo.co.in (SVA)}, Burkert, A$^{4},^{5}$\thanks{Max-Planck Fellow} and Kuiper, R$^{2}$ \\
$^{1}$Indian Institute of Technology, Kharagpur, West Bengal, India\\
  $^{2}$ Institute of Astronomy and Astrophysics, University of T$\ddot{\mathrm{u}}$bingen, Auf der Morgenstelle 10, D-72076 T$\ddot{\mathrm{u}}$bingen, Germany\\
  $^{3}$ Harvard-Smithsonian Center for Astrophysics, 60 Garden Street, Cambridge, MA 02138, USA \\
$^{4}$ University Observatory Munich, Schneirstrasse 1, D-81679, Munich, Germany \\
$^{5}$ Max Planck Institut f$\ddot{u}$r extraterrestrische Physik, Geissenbachstrasse, D-85748, Garching, Germany \\}
\begin{document}

\date{Accepted 0000 December 00. Received 0000 December 00; in original form 0000 October 00}

\pagerange{\pageref{firstpage}--\pageref{lastpage}} \pubyear{2002}

\maketitle

\label{firstpage}

\begin{abstract}
  The star-forming ability of a molecular cloud depends on the fraction of gas it can cycle into the dense-phase. Consequently, one of the crucial questions in reconciling star-formation in clouds is to understand the factors that control this process. While it is widely accepted that the variation in ambient conditions can alter significantly the ability of a cloud to spawn stars, the observed variation in the star-formation rate in nearby clouds that experience similar ambient conditions, presents an interesting question. In this work we attempted to reconcile this variation within the paradigm of colliding flows. To this end we develop self-gravitating, hydrodynamic realisations of identical flows, but allowed to collide off-centre. Typical observational diagnostics such as the gas-velocity dispersion, the fraction of dense-gas, the column density distribution ({\small N-PDF}), the distribution of gas mass as a function of $K$-band extinction and the strength of compressional/solenoidal modes in  the post-collision cloud were deduced for different choices of the impact parameter of collision. We find that a strongly sheared cloud is terribly inefficient in cycling gas into the dense phase and that such a cloud can possibly reconcile the sluggish nature of star-formation reported for some clouds. Within the paradigm of cloud-formation via colliding flows this is possible in case of flows colliding with a relatively large impact parameter. We conclude that compressional modes -  though probably essential -  are insufficient to ensure a relatively higher star-formation efficiency in a cloud.
\end{abstract}


\section{Introduction}
It is well-known that star-formation in MCs occurs in the densest pockets of gas within these clouds. Consequently, a theory attempting to explain the formation of stars must also reconcile the possible impact of ambient conditions on the ability of a cloud to cycle gas into these putative star-forming pockets. Numerous recent observational and numerical efforts have been dedicated to this cause.\\ \\
In this connection attention may be drawn to the following interesting results :
(\textbf{1}) Battisti \& Heyer (2014) showed that the dense gas fraction i.e., the fraction of gas having density typically greater than $\sim 10^{3}$ cm$^{-3}$, remained more or less constant as a function of the Galactic radius. (\textbf{2}) On the contrary, Kruijssen \emph{et al.} (2014) showed that gas in the Galactic {\small CMZ} is largely stable against the gravitational instability from which it may be inferred that the fraction of the putative star-forming gas (the so-called star-forming fraction ({\small SFF}); this is typically the fraction of gas having column density in excess of $\sim 10^{21}$ cm$^{-2}$), must decline towards the Galactic centre, despite its predominantly molecular composition. This is in fact true as star-formation in the {\small CMZ} is known to be quite inefficient (e.g. Eden \emph{et al.} 2012, Eden \emph{et al.} 2013, Longmore \emph{et al.} 2014). (\textbf{3}) Ragan \emph{et al.} (2016) showed that the {\small SFF} in the Galactic disk declines steadily with increasing Galactic radius. \\ \\
Recent work by Roman-Duval \emph{et al.} (2010), Koda, Scoville \& Heyer (2016) and Ragan \emph{et al.} (2016) not only reinforces the fact that the fraction of putative star-forming gas tends to decline towards the Galactic centre, but shows that this fraction also declines at relatively large Galactic radii. This leads to the inference that clouds in the Solar neighbourhood must somehow exhibit a greater propensity towards star-formation. However, that is hardly true. For instance, Lada \emph{et al.} (2009; 2010) showed that {\small MC}s even within the Solar neighbourhood exhibited a significant variation in their ability to form stars, or for that matter, in the total gas-mass available at large extinction. Ragan \emph{et al.} (2016) argued that these observed variations could not possibly be reconciled by merely attributing them to the difference in ambient environment of individual clouds. \\ \\
Indeed, this is especially the case with clouds in the Solar neighbourhood where the clouds in question experience mutually similar ambient conditions. Consequently, Ragan \emph{et al.} suggest that the star-forming ability of a cloud might depend crucially on the physical properties inherited by it during its formation. In a numerical investigation of the problem of formation of {\small MC}s in a galactic disk Dobbs \emph{et al.} (2012) demonstrated cloud-formation due to the merger of gas-flows in the disk superposed with a spiral-shock. Similarly, Bonnell \emph{et al.} (2013) demonstrated that a Kennicutt-Schmidt type star-formation law could be reproduced in a self-gravitating realisation of a galactic-disk where the interstellar medium evolves via the interplay between the thermal and gravitational instability. There is another genre of models in which star-formation is largely turbulence-regulated with/out the magnetic field. According to this latter model the observed variations in the star-formation efficiency ({\small SFE}), i.e. the fraction of gas that actually ends up in stars, across clouds could possibly be reconciled if the turbulent velocity field in these clouds were dominated more by solenoidal modes rather than compressional modes (e.g. Federrath \emph{et al.} 2010; Kainulainen \emph{et al.} 2014).  In a more recent contribution Federrath \emph{et al.} (2016) also argued that the observed relatively low star-formation rate in the \emph{Brick} located in the Galactic Central Molecular Zone is probably due to the dominance of the solenoidal modes of the turbulent velocity field and likely induced by large-scale shear. \\ \\
Despite the success of global cloud-formation models of the kind described above in recovering the Schmidt-Kennicutt type star-formation law, it has not been possible for them to reconcile the observed variations in cloud-properties. The exorbitant computational costs incurred in developing such simulations is an obvious limitation. Similarly, the success of the turbulence-regulated star-formation model is largely based on the choice of the strength and the nature (compressional \emph{vs} solenoidal) of the initial turbulent velocity field. Although in a recent contribution (Anathpindika \emph{et al.} 2017), hereafter referred to as paper I, we demonstrated the variation in various cloud-properties as a function of interstellar pressure. In this sequel we wish to investigate if it is possible to reconcile variations in the initial conditions for star-formation in a {\small MC} for a given magnitude of external pressure, especially for the magnitude of pressure typically estimated in the the Solar neighbourhood. Equivalently speaking, we propose to investigate if for a given magnitude of interstellar pressure the {\small SFF}, or indeed the distribution of column density in the assembled cloud can vary.\\ \\
 We also investigate if it would be possible to assemble a cloud dominated by solenoidal modes as required by the turbulence-regulated star-formation model. To this end we will investigate the {\small SFF}, the dense gas fraction (the fraction of gas having density in excess of $\sim 10^{3}$ cm$^{-3}$), the distribution of column density ({\small N-PDF}), for gas in the assembled cloud, the distribution of gas mass as a function of visual extinction and calculate the relative strength of the compressional and solenoidal modes in the assembled cloud. We emphasise, it is not our extant interest to resolve individual prestellar cores in these simulations. We leave the investigation of formation and fragmentation of density structure within clouds to a later work. The layout of the paper is follows. The set-up of our simulations and the numerical algorithm used will be described in \S 2. Thereafter, results from our realisations will be presented and discussed in \S 3 and 4 respectively. We will conclude in \S 5.
\begin{figure}
\label{Figure 1}
\vspace{1pc}
\centering
\includegraphics[angle=270,width=0.5\textwidth]{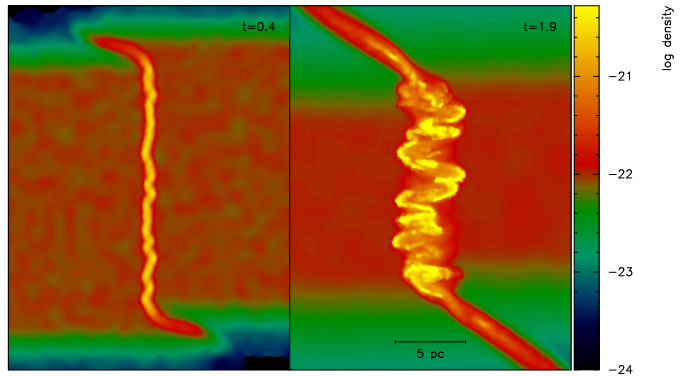}
\caption{Rendered density (gas density in units of g cm$^{-3}$) images of the cross-section through the plane of collision of identical cylindrical flows. The impact parameter of the collision was b$_{impact}$=0.2*R$_{inf}$. \emph{Left-hand panel} : As usual, the post-collision slab is attended by the Non-linear thin shell instability as evidenced by the appearance of corrugations on it. \emph{Right-hand panel} : With time, growth of the shell instability quickly saturates causing the slab to buckle. The epoch in units of Myr is shown in the top right-hand corner of each figure.}
\end{figure}
%----------------------------------------------------------
\section{Initial Conditions and Numerical Method}
We use the same set-up as the one described in paper I where flows having initially uniform density were allowed to collide head-on. Here we use the initial conditions of realisation  listed 6 in that work, but now allow the flows to collide at a finite impact parameter, $b_{impact}$. Formation of clouds via collision between gas-flows has been widely explored and the interested reader may refer \S 1 of paper I for a brief survey of the corresponding literature. Collision between gas-streams can possibly be reconciled within the paradigm of cloud-formation due to the propagation of a density-wave in the disk of a galaxy. In this scenario clouds are assembled in the crests and troughs of the density wave where gas-flows converge at random orientations relative to each other. \\ \\
A more formal treatment of cloud-formation should include many more details - for a start, the flows assembling a cloud should be warm-neutral which makes it essential to have a recipe to convert atomic hydrogen to molecular. Presently we do not have one and so, as in paper I, the implementation of gas-cooling was done with the aid of a cooling function mimicking the cooling curve for the \small{ISM}. However, even then it is difficult to envisage a situation where the initially diffuse neutral gas can be assembled into a cloud of molecular gas without invoking some sort of a mechanism to converge the flows. To circumvent this conundrum we chose to prefer the observational findings of Kuno \& Nakai (1997), who reported evidence for gas streamlines in the disk of the {\small M51} with CO-mapping of the disk, to motivate the colliding flows set-up. Studies of gas-flows in the {\small M81}, another grand-design spiral, by Visser (1980), showed they were large consistent with predictions of the density wave theory. The general idea then being that large streams of gas were compressed as they entered the spiral density perturbations in the disk of a galaxy. Since CO is an excellent tracer of molecular hydrogen the work by the former authors may also be taken as evidence to support the set-up such as the one used in this work where the pre-collision flows are somewhat denser than typical neutral gas-flows. That being so, we discuss some caveats associated with this set-up in \S 4.1 below.\\ \\
The physical details of the pre-collision cylindrical inflows are - $L_{inf}$ = 50 pc, $R_{inf}$ = 10 pc, $V_{inf}$=10.36 km/s, $\bar{n}_{gas}$ = 50 cm$^{-3}$, $T_{gas}$ =  500 K, where $L_{inf}$, $R_{inf}$, $V_{inf}$, $\bar{n}_{gas}$ and $T_{gas}$ are respectively the length, radius, magnitude of inflow-velocity, average density and the temperature of the gas constituting the respective inflows. The magnitude of external pressure corresponding to our choice of inflow-velocity is, $P_{ext}/k_{B}\sim$ 6.51$\times 10^{5}$ K cm$^{-3}$, which is consistent with the estimates for interstellar pressure in the Solar neighbourhood (e.g. Kasparova \& Zasov 2008). We appreciate that gas-flows in a typical galaxy would  undoubtedly be much longer than the restrictive dimensions we have selected here since the limited computational resources at our disposal place an obvious limit. So we would rather look at these flows as chunks/packets within individual, extended gas-streamlines in the disk of a typical galaxy. This is consistent with the characteristics of gas-flow deduced for streams in the arms of the {\small M51} (e.g. Kuno \& Nakai 1997). Thus despite the relative simplicity and the idealised nature of our set-up, exploiting it to study the assembly of a cloud in light of the available observational deductions about gas-flows in the disk of a typical spiral galaxy, may not be totally inappropriate. \\ \\
Ideally we should be performing a large number of realisations with a random choice of the impact parameter, $b_{impact}$, but such an exercise would be computationally unfeasible. And so, we perform 4 realisations that broadly cover possible scenarios of gas-flow orientation.
Simulations were developed for four choices of impact parameter, viz., $b_{impact}$ = 0.01 $R_{inf}$, 0.2 $R_{inf}$, 0.5 $R_{inf}$ and 1.8 $R_{inf}$. The first of these realisations is almost akin to the case of a head-on collision while the last one is where the in-flows merely graze past each other. As in paper I, the hydrodynamic simulations developed in this work are self-gravitating. The gas in the inflows is assumed to be composed of the usual cosmic mixture having mean atomic weight, $\mu$ = 2.29. Simulations were performed using the {\small SPH} code {\small SEREN} (e.g. Hubber \emph{et al.} 2011). Finally, the numerical details such as the spatial resolution of the realisations and the implementation of dynamical cooling with the aid of a cooling curve remain as in paper I. Realisations in this work were developed with $N_{gas}\sim$ 2.16$\times 10^{6}$ number of gas particles.
%--------------------------------------------------------------------
\begin{figure*}
\label{Figure 2}
\vspace{1pc}
\centering
\includegraphics[angle=270,width=\textwidth]{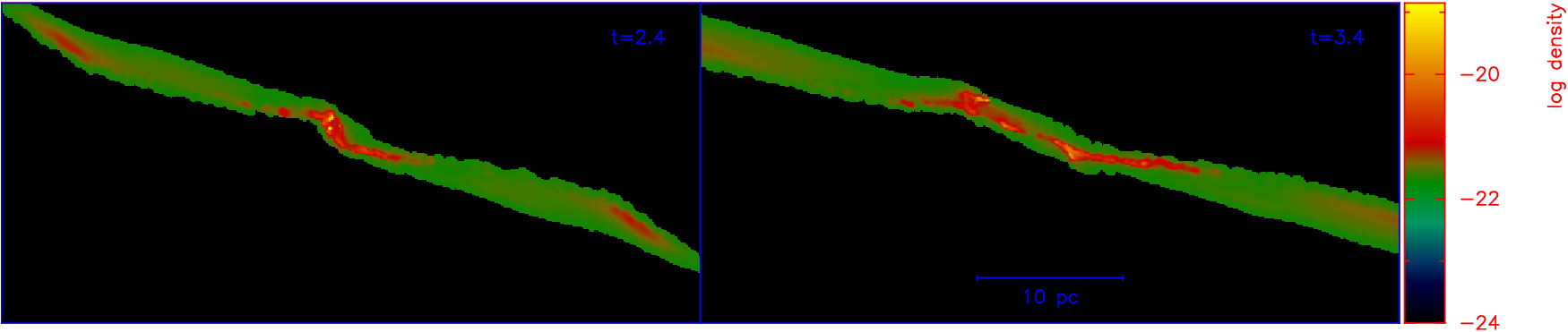}
\includegraphics[angle=270,width=\textwidth]{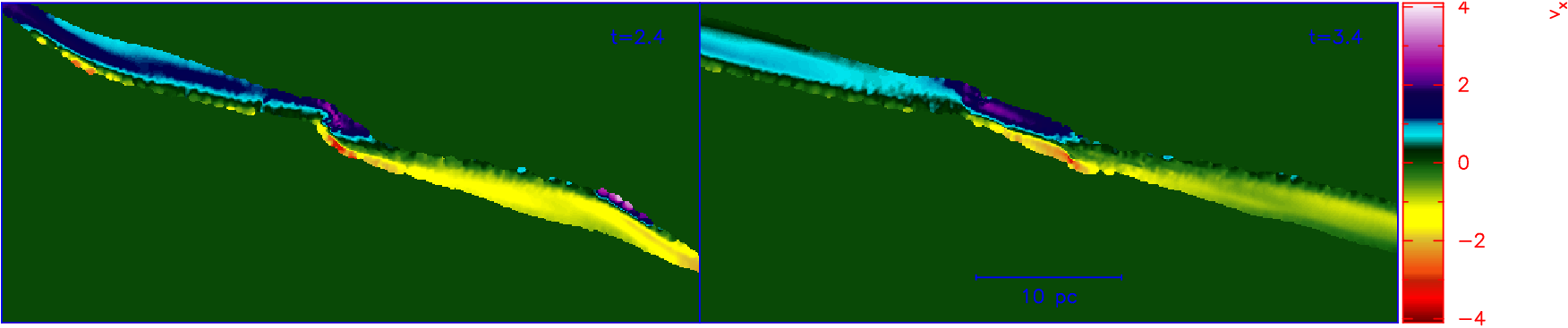}
\caption{\emph{Upper-panel} : As in Fig.1, but now, b$_{impact}$=1.8*R$_{inf}$, so that the post-shock slab is strongly sheared which results in the upper and lower lobes. \emph{Lower-panel} : This is a rendered image of the position-velocity profile for the post-collision body. The strong shear naturally induces a velocity gradient within each lobe of the post-collision body; velocity in units of km/s. }
\end{figure*}
%-----------------------------------------------------------------
\section{Results}
\subsection{Evolution of the shocked slab}
Fragmentation of the post-collision slab has been examined by a number of authors (e.g. V{\' a}zquez-Semadeni \emph{et al.} 2007; Heitsch \emph{et al.} 2008; Anathpindika \emph{et al.} 2009 a and Carroll-Nellenback \emph{et al.} 2014). As a result, it is now well established that the post-collision shocked slab evolves via the interplay between the non-linear thin shell instability ({\small NTSI}) and the thermal instability. In an earlier work Anathpindika (2009) examined numerically the evolution of an oblique shocked-slab whereas the stability of such slabs against the gravitational instability was analytically examined by Usami \emph{et al.} (1995). It has been argued by these authors that the oblique shocked-slab being largely dominated  by shearing motion between slab-layers is stable against the {\small NTSI}  and the gravitational instability. The shocked slab in this case has three distinct regions viz. the upper and lower lobe and the central planar region, the spatial extent of which diminishes with increasing impact parameter, $b_{impact}$. \\ \\
Shown on the upper panel of Fig. 1 is the rendered density image of the post-collision slab for the realisation with $b_{impact}$ = 0.2 $R_{inf}$ at two epochs. With a relatively small impact parameter, the post-collision slab is largely planar with small lobes on the top and bottom. As was the case with planar shocked-slabs discussed in Paper I, the {\small NTSI} is triggered soon after the slab is assembled. Thereafter, this instability grows rapidly and causes the slab to buckle as is visible on the left-hand panel of Fig. 1, and eventually the slab becomes bloated as the instability saturates; see picture on the right-hand panel of Fig. 1. By contrast, collisions with a larger impact parameter induce a stronger shearing interaction between gas-layers within the post-collision slab which tends to stabilise the post-collision slab against various dynamical instabilities. \\ \\
In a dynamical set-up such as the one where gas-streams/clouds are allowed to collide, the relevant timescale is the so-called cloud-crushing time ($t_{cr}\sim \frac{2L_{inf}}{V_{inf}}$), which is the timescale on which the colliding clouds/flows are consumed by the post-collision shock-front. However, this timescale, or indeed the free-fall time of the pre-collision gas,  turns out to be significantly longer than the growth-time of the {\small NTSI}. Anathpindika (2009) showed that the {\small NTSI} growth-timescale is only a fraction of the cloud-crushing time and that a slab resulting from the off-centre collision of clouds was stable against the {\small NTSI}. In the first three realisations of the present work, where the impact parameter of the collision is relatively small and where the {\small NTSI} appears to play a significant role, the post-collision slab became bloated as the growth of the {\small NTSI} saturated; the calculations in these respective cases were terminated soon after. On the other hand, calculations in realisation 4 where the collision was highly off-centre such that the flows  merely grazed past each other, there was no evidence of the {\small NTSI} as was also suggested by Anathpindika (2009). This calculation was terminated when the respective inflows crossed over to the rear-end of the other.\\ \\
While the observed evolution of the slab in this work and that reported earlier by Anathpindika (2009) is qualitatively similar, the scope of investigation in the latter was restricted. For a start, the initial conditions in that work, individual Bonnor-Ebert spheres, represented at best small, turbulent clumps of gas in a {\small MC}. Depending on the pre-collision velocity of the clumps, the collision led either to a larger clump due to coalescence or a slab that evolved via the interplay between the self-gravity of the slab and the {\small NTSI}. On the other hand, physical details of the in-flows in this work are significantly different to the initial conditions in the former. The slab assembled in the present set of simulations is approximately 2 orders of magnitude more massive that that assembled by the colliding clumps in the former work. And so despite the qualitative similarity in evolution of the post-collision slab, the significantly stronger self-gravity in this case must influence the gas-diagnostics in it. \\ \\
Shown on the upper left and right-hand panel of Fig. 2 is the slab for the realisation where the pre-collision impact parameter, b$_{impact}\sim$ 1.8$R_{inf}$, was relatively large. As the inflows grazed past each other they produced spectacular lobes. The corrugations on the post-collision slab, usually associated with the shell-instability, are noticeably absent in this case as is evident from the plots on these panels of Fig. 2. 
%=================================================================
%------------------------------------------------------------------------
\begin{figure}
\label{Figure 3}
\vspace{1pc}
\centering
\includegraphics[angle=270,width=0.5\textwidth]{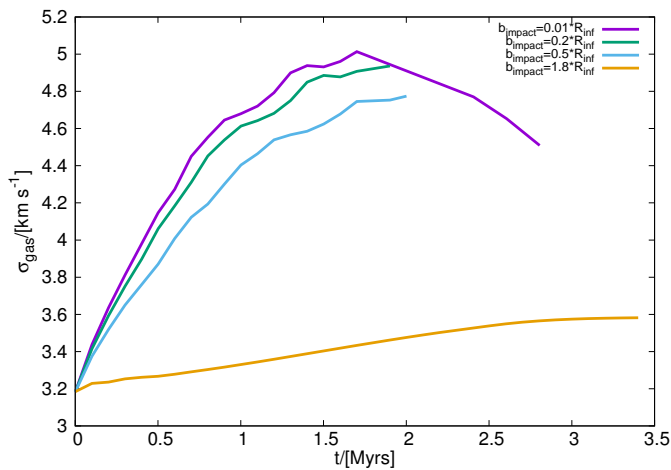}
\caption{Temporal variation of the velocity dispersion induced in the post-collision slab for each choice of the impact parameter, b$_{impact}$.}
\end{figure}
Shown on the lower-panels of this figure are the rendered velocity-plots corresponding to the respective epochs. The presence of a velocity gradient typically on the order of a km s$^{-1}$ pc$^{-1}$ along each lobe is visible from these plots. This is also consistent with the velocity gradient reported by Lada \emph{et al.} (2009) in the California {\small MC}.
\subsection{Diagnostics of gas in the post-collision slab}
\textbf{Gas-velocity dispersion}\\
The temporal evolution of the gas velocity dispersion, $\sigma_{gas}$, in the post-collision slab for different choices of the impact parameter is shown in Fig. 3. As discussed above, the {\small NTSI} dominates the slab evolution in the first three realisations where the impact parameter, $b_{impact}$, of the collision is relatively small. In these cases the variation of $\sigma_{gas}$ is qualitatively similar to that reported in paper I for flows colliding head-on. In all these realisations the velocity dispersion, $\sigma_{gas}$, acquires a maxima before eventually petering-out. We also note here that the peak magnitude of $\sigma_{gas}$ decreases progressively with increasing $b_{impact}$, or equivalently, the increasing obliqueness of the post-collision shock tends to stabilise the shocked-slab against the {\small NTSI} that limits the magnitude of $\sigma_{gas}$. Consequently, the slab is most stable for the largest impact parameter where the magnitude of $\sigma_{gas}$ is also the smallest. Also, in this latter case the phase where $\sigma_{gas}$ declines is not visible. Now, although a relatively lower magnitude of $\sigma_{gas}$ is likely to assist star-formation, we shall have to see if indeed the large impact parameter of collision also cycles more gas into the dense-phase.\\\\
%--------------------------------------------------------------------
\begin{figure*}
\label{Figure 4}  
\vspace{1pc}
\centering
\includegraphics[angle=270,width=0.45\textwidth]{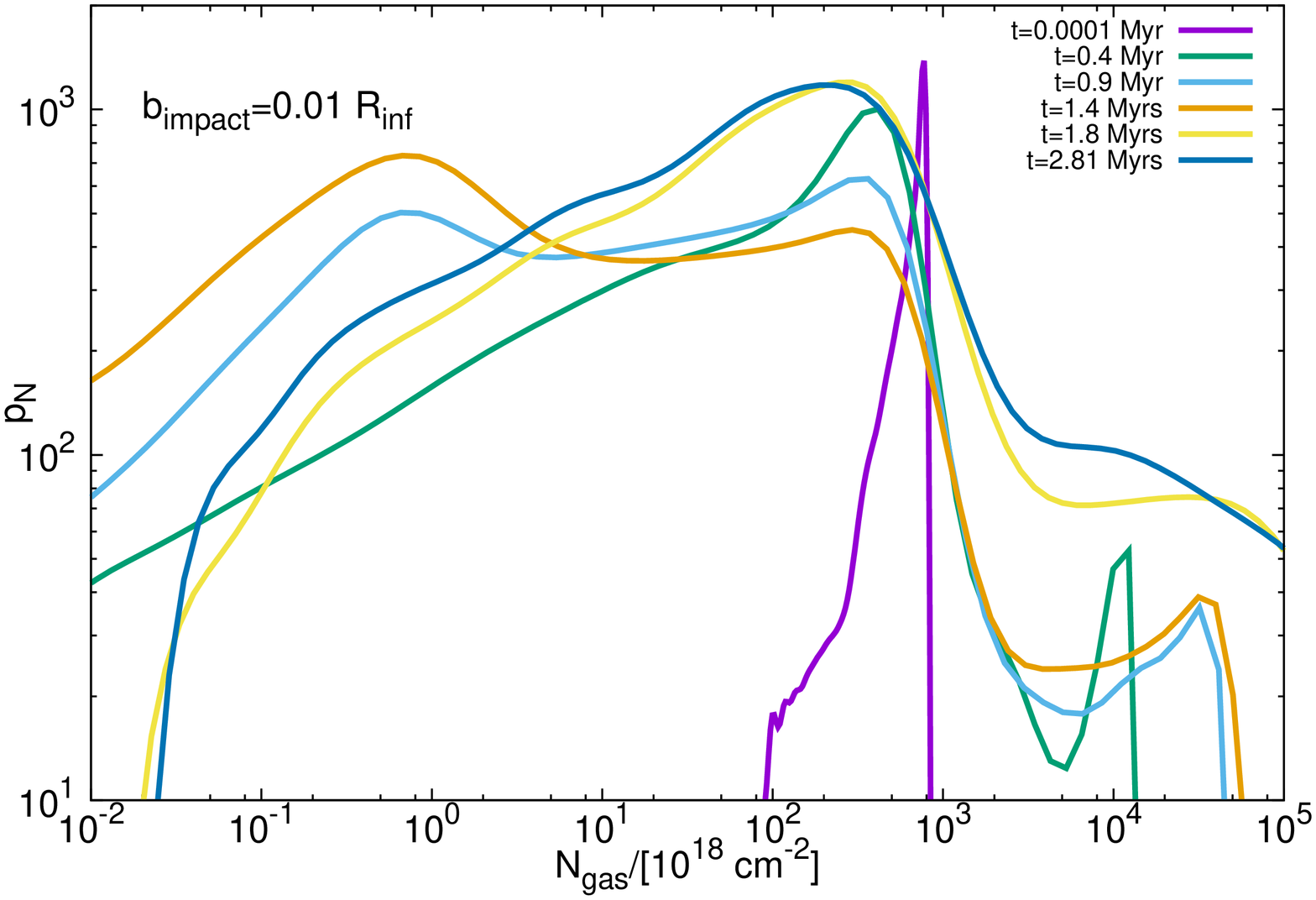}
\includegraphics[angle=270,width=0.45\textwidth]{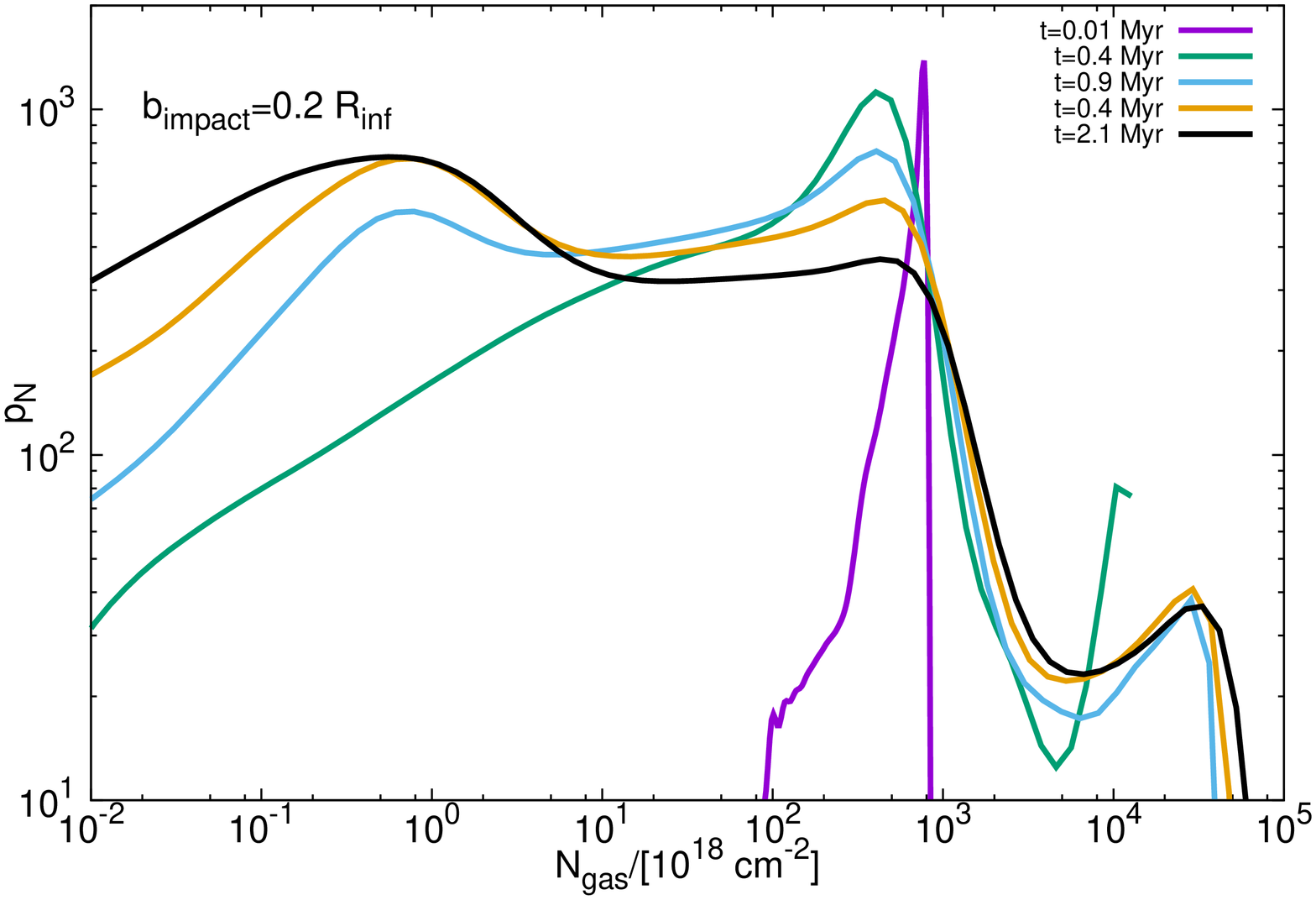}
\includegraphics[angle=270,width=0.45\textwidth]{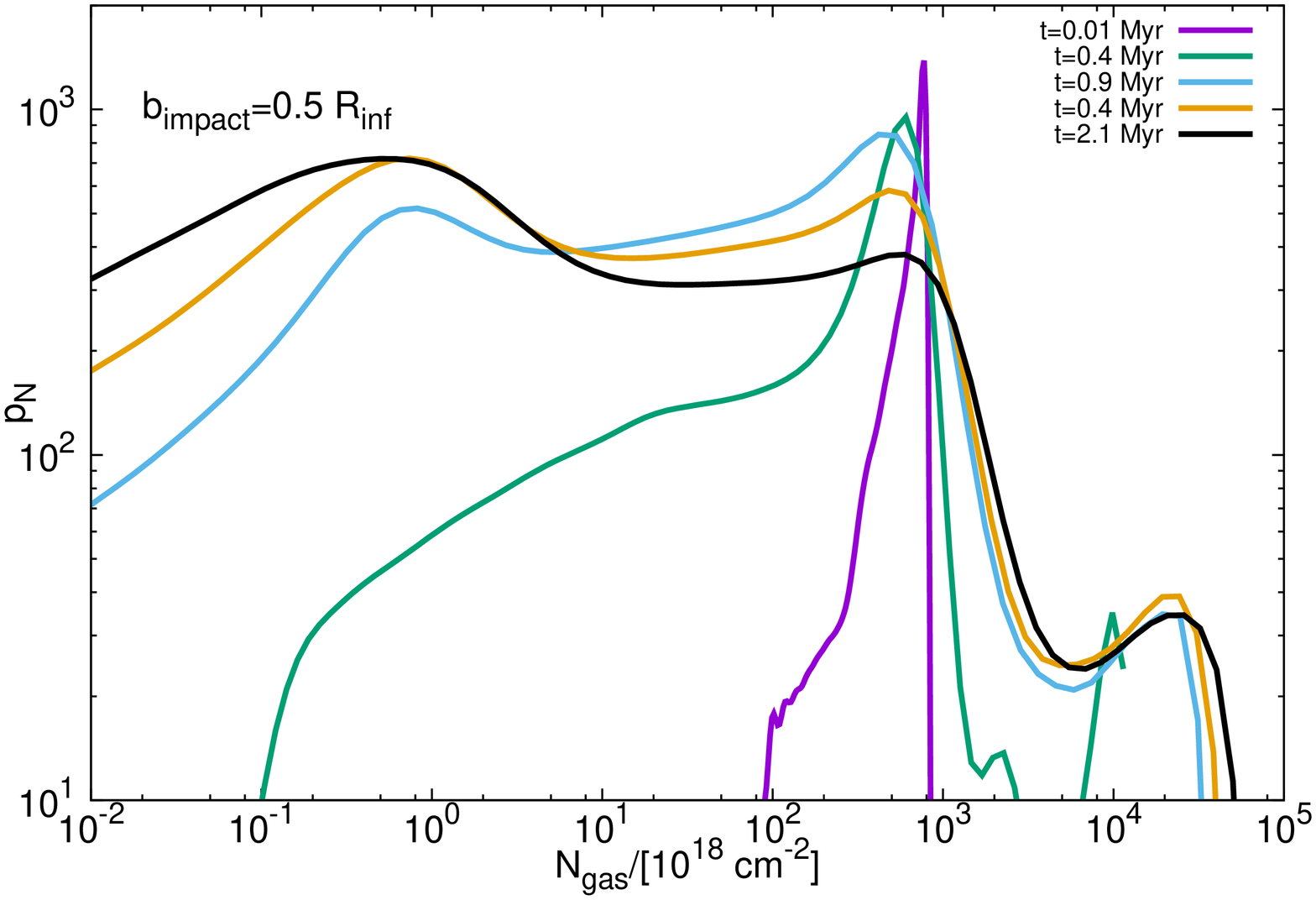}
\includegraphics[angle=270,width=0.45\textwidth]{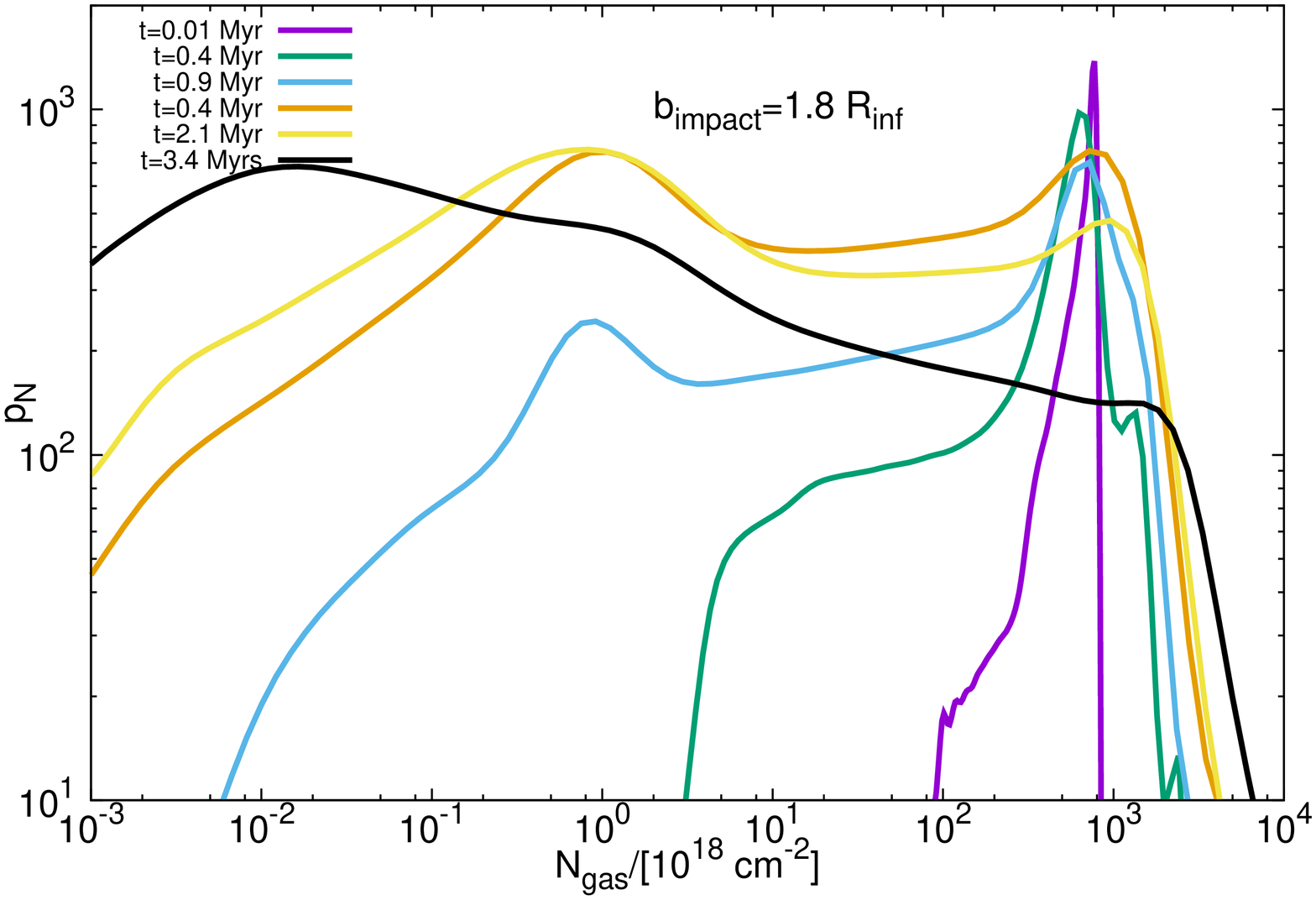}
\caption{{\small N-PDF}s for respective choices of the impact parameter, $b_{impact}$. Notice the steepening of the {\small N-PDF}s at the high-density end with increasing magnitude of the impact parameter.}
\end{figure*}
%-------------------------------------------------------------------
\textbf{Column density probability distribution function ({\small N-PDF})}\\
The column density of gas in the slab was calculated in the plane of the post-collision slab i.e., orthogonal to the direction of the respective inflows. For the purpose, gas particles in the post-collision slab were distributed on an uniform grid of size 256$^{3}$. The column density on each cell was calculated by summing over weighted contributions from individual gas particles to the respective cell; the standard cubic-spline was used as a weighting function.
Shown on the respective panels of Fig. 4 are the column density probability distribution function ({\small N-PDF}) at different epochs for each choice of $b_{impact}$. The {\small N-PDF}s for a relatively small impact parameter are similar to those reported in paper I for a head-on collision between inflows. Soon after the slab is assembled, the respective {\small N-PDF}s, especially those in the first three realisations, do not exhibit much extension into the high-density regime. Thereafter, as it continues to accrete gas from the inflows, the {\small N-PDF}s develop a power-law tail at the high-density end at later epochs. As pointed out in the introduction, it is not our extant interest to follow the evolution up to the point of core-formation and so the power-law tail at the high- density is not extensive. The objective here is only to draw attention to the deviation of the {\small N-PDF} from lognormal distribution and the fraction of gas that ends in the dense regime. \\ \\
The following features are readily visible from these plots :
first, it is evident that {\small N-PDF}s having a variety of shapes can be generated by merely varying the impact parameter of collision. Second, the {\small N-PDF} in the bottom right-hand panel reinforces the point that the collision in this case (grazing incidence - realisation 4), is inefficient in cycling gas to relatively high column densities. This point can be better appreciated if we only look at the {\small N-PDF}s just before the respective realisations were terminated. These have been shown in  Fig. 5 and overlaid on these {\small N-PDF}s are power-law and lognormal fits shown respectively with dashed and continuous lines. It clear that irrespective of the choice of the impact parameter the extension of the {\small N-PDF} at the high-density end is not always a single power-law. Furthermore, this power-law tail at the high-density end exhibits a break and sometimes the extension after this break, as in realisations 2 and 3, appears log-normal. On the contrary, in realisations 1 and 4 this extension is a power-law though the slope of the {\small N-PDF} for the latter realisation is significantly steeper than that in the former. Finally, a feature common to all the {\small N-PDF}s shown in this figure is the approximately log-normal composition of the respective {\small N-PDF}s at lower column densities. For instance, this combination of log-normal fits is shown using a continuous line in case of the {\small N-PDF} for realisations 2 and 3. \\ \\
\begin{figure}
\label{Figure 5}  
\vspace{1pc}
\centering
\includegraphics[angle=270,width=0.5\textwidth]{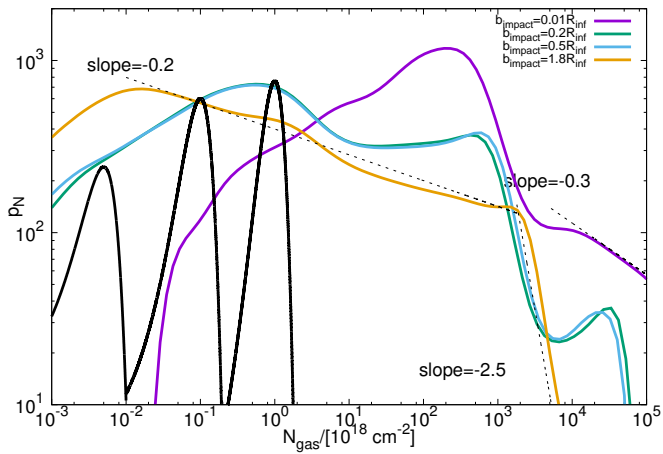}
\caption{The {\small N-PDF}s for the assembled cloud in the respective realisations just before they were terminated. The dotted lines represent the power-law fits to different portions of the individual {\small N-PDF}s, while the log-normal distributions at the low-density end are shown by a continuous line. }
\end{figure}
Physically speaking, the shape of the respective {\small N-PDF}s may be reconciled as follows : In the first three realisations where the {\small NTSI} played a dominant role in the evolution of the post-collision slab, the appearance of the log-normal extension at the low-density end is attributed to the {\small NTSI}-induced bloating of the slab as discussed in \S 3.1. On the other hand, at the other extreme, we have seen that the realisation in which the inflows merely grazed past each other, and where the {\small NTSI} remained subdued, was most inefficient in cycling gas into the dense-phase. Consequently, much of the gas in this case remained in the low-density regime. We also wish to point that the inflows themselves could not possibly have expanded before collision, for the expansion timescale of the individual in-flow, $t_{exp}$, compares with the crushing time, $t_{cr}$, defined above, as $\frac{t_{exp}}{t_{cr}}\propto \frac{V_{inf}}{a_{0}}$; $a_{0}$, being the sound-speed in the pre-collision flow and which is identical for individual flows. This ratio is significantly larger than unity implying that the pre-collision flows are unlikely to expand before they collide.\\ \\ 
While it may be argued that the tail at the high-density end of the {\small N-PDF} for realisations 2, 3 and 4 could itself be approximated by another lognormal instead of a steep power-law as shown in Fig. 5, it hardly alters the inference that the fraction of gas cycled into the dense-phase decreases significantly with increasing impact parameter. In fact, the steepening of the slope of the power-law is only another way of suggesting that the tail is clipped off, making way to a lognormal-like distribution. The contrast between the relatively shallow tail for realisation 1 (slope=-0.3), and that for the rest is remarkable. Interestingly, the distribution at the high-density end even develops a log-normal bump for realisations 2 and 3. Similar {\small N-PDF}s have been reported for some regions in the cloud Mon {\small R2} (e.g. Pokhrel \emph{et al.} 2016). The {\small N-PDF} for a number of {\small MC}s in the Solar neighbourhood were deduced by Lombardi \emph{et al.} (2015). \\ \\
The relatively steep tail at the high-density end of the {\small N-PDF} deduced for the post-collision slab in realisations 2, 3 and 4 is similar to the fits provided by Lombardi \emph{et al.} to the {\small N-PDF}s deduced by them for clouds such as the Polaris, Pipe, California and the Perseus that show little star-forming activity. Also, the low-density end of the {\small N-PDF}s for the said realisations shown in Fig. 5 is qualitatively similar to that deduced by Lombardi \emph{et al.} for these clouds in the sense that a significant fraction of gas in these clouds remains at low-extinction, perhaps below the detection threshold of the survey, and so the respective {\small N-PDF}s are incomplete at small extinction.
On the other hand, the {\small N-PDF} deduced in realisation 1 appears similar to those deduced for star-forming clouds such as the Orion, Ophiuchus and the Taurus. Evidently, not only the ambient conditions i.e., the magnitude of interstellar pressure, as argued in paper I, affect the gas-distribution, but also the impact parameter of collision significantly alters the nature of the {\small N-PDF} and therefore, the fraction of putative star-forming gas.\\ \\
%---------------------------------------------------------------------------
\textbf{Dense gas-fraction and the star-forming fraction}\\
As in paper I we define the dense gas-fraction, M$_{frac}$, as the fraction of gas having density, $n \gtrsim $ 10$^{3}$ cm$^{-3}$ and being colder than $\sim$50 K. Although somewhat ad hoc, this density and temperature threshold is good enough to estimate the fraction of gas likely to end up in putative star-forming clumps (e.g. Ragan \emph{et al.} 2016 and other references therein). Furthermore, we also use a higher density threshold, $n \gtrsim 10^{4}$ cm$^{-3}$, to trace the fraction of strongly self-gravitating gas in a pristine cloud. In typical star-forming clouds such dense pockets of gas are traced due to emission from molecules such as {\small HCN}, {\small HNC} and {\small HCO$^{+}$} (e.g. Gao \& Solomon 2004, Usero \emph{et al.} 2015 and Biegel \emph{et al.} 2016). \\ \\
The dense gas-fraction, M$_{frac}$ $\equiv\frac{M_{thresh}(n \gtrsim 10^{3} \mathrm{cm}^{-3}, 10^{4} \mathrm{cm}^{-3}; 50 \mathrm{K})}{M_{gas}}$, for each choice of $b_{impact}$ is shown on the upper panel of Fig. 6. Irrespective of the choice of the impact parameter and the density-threshold, $n$, the fraction, M$_{frac}$, increases steadily with time, but at best cycled a little under $\sim 10\%$ of the gas into the dense-phase at the time calculations were terminated. Indeed, this fraction is significantly smaller in the realisation where the inflows merely grazed past each other. While the fraction of gas cycled into the dense-phase could possibly increase as the post-collision slab continues to accrete gas from the colliding flows, it is unlikely to exhibit a dramatic increase. This is because the {\small NTSI} destabilises the shocked-slab and causes fluid-layers to mix, thereby rupturing the dense pockets of gas. In any case it is evident that grazing inflows are most inefficient in cycling gas into the dense-phase. The fraction of gas cycled to densities higher than $\sim 10^{4}$ cm$^{-3}$ is typically an order of magnitude smaller. Evidently, star-formation in a pristine cloud must be quite inefficient. We also note, irrespective of the choice of the density threshold and the impact parameter the characteristics of the dense gas-fraction look mutually similar. This is because whatever the choice of the pre-collision impact parameter, the fraction of gas cycled into the dense-phase is limited by shearing interaction between layers of gas in the shocked slab. The source of shear, however, is different in the extreme scenarios of near head-on collision and one in which the gas-flows merely graze past each other. In the the former shear is induced by the growth of the {\small NTSI} while in the latter, it is the obliqueness of the collision that induces a strong shearing interaction between layers of gas. \\ \\
Alternatively, the fraction of putative star-forming gas may also be quantified using a column density threshold which as in paper I is fixed at $\sim 10^{21}$ cm$^{-2}$. We refer to this fraction as the star-forming fraction ({\small SFF}); see plot on the lower-panel of Fig. 6. As with the dense gas-fraction, the {\small SFF} also rises in incremental fashion, but contrary to the former, the {\small SFF} is the highest for the realisation with the largest b$_{impact}$. The reason being the relatively greater stability of the post-collision slab with higher impact parameter against dynamic instabilities. The {\small SFF} peters-out at later epochs for the case of grazing-incidence as the post-collision slab is strongly sheared. On the other hand, for relatively smaller b$_{impact}$, the {\small SFF} is the lowest for the case where the collision was almost head-on, i.e. where b$_{impact}$ was the smallest, and consequently, the {\small NTSI} growth must be the strongest. At any rate the {\small SFF} remains typically less than a percent which could possibly help us reconcile the observationally reported poor efficiency of star-formation in nearby {\small MCs}. We will revisit this point later in \S 4.\\ \\
%------------------------------------------------------------
\begin{figure}
\label{Figure 6}  
\vspace{1pc}
\centering
\includegraphics[angle=270,width=0.5\textwidth]{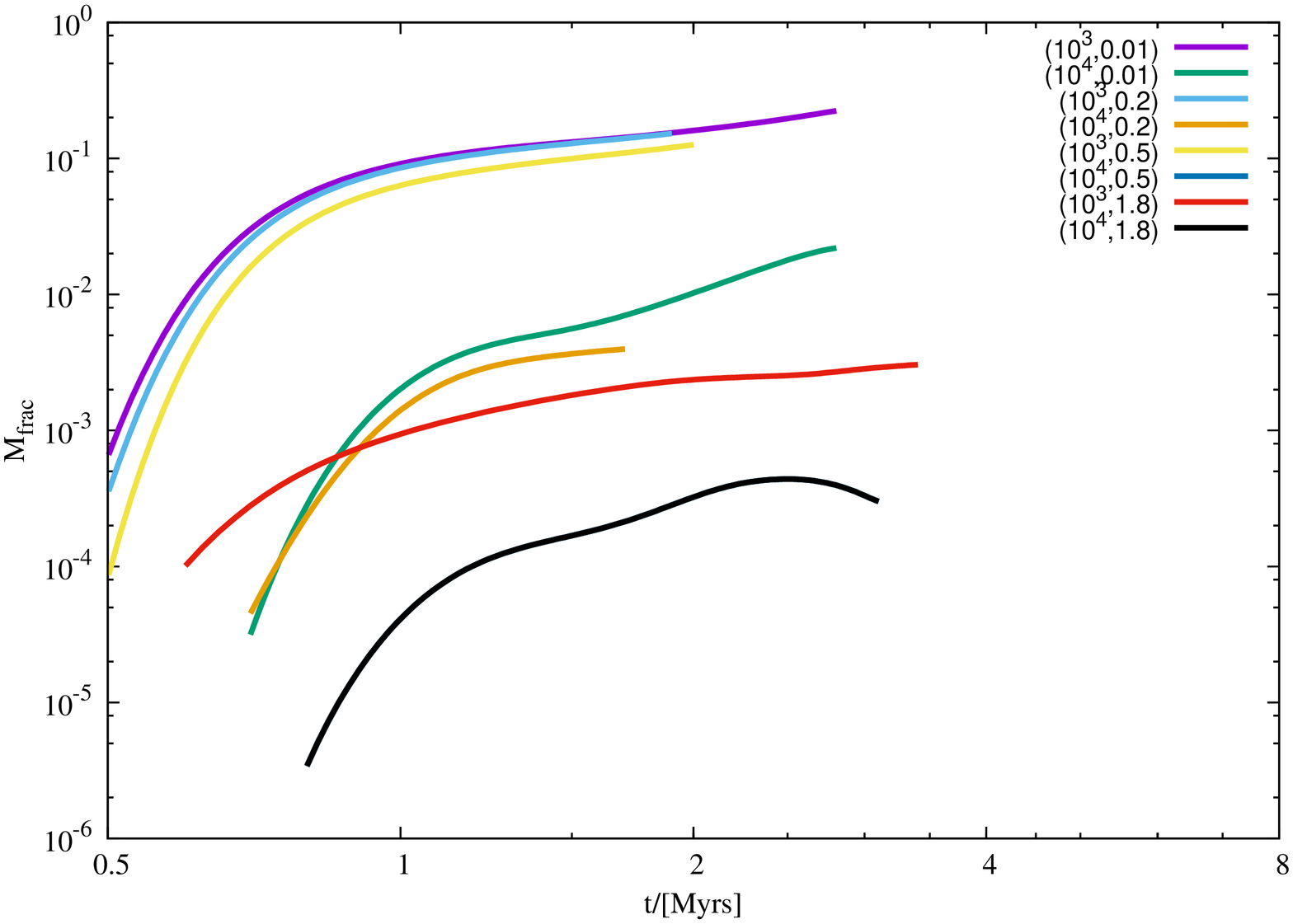}
\includegraphics[angle=270,width=0.5\textwidth]{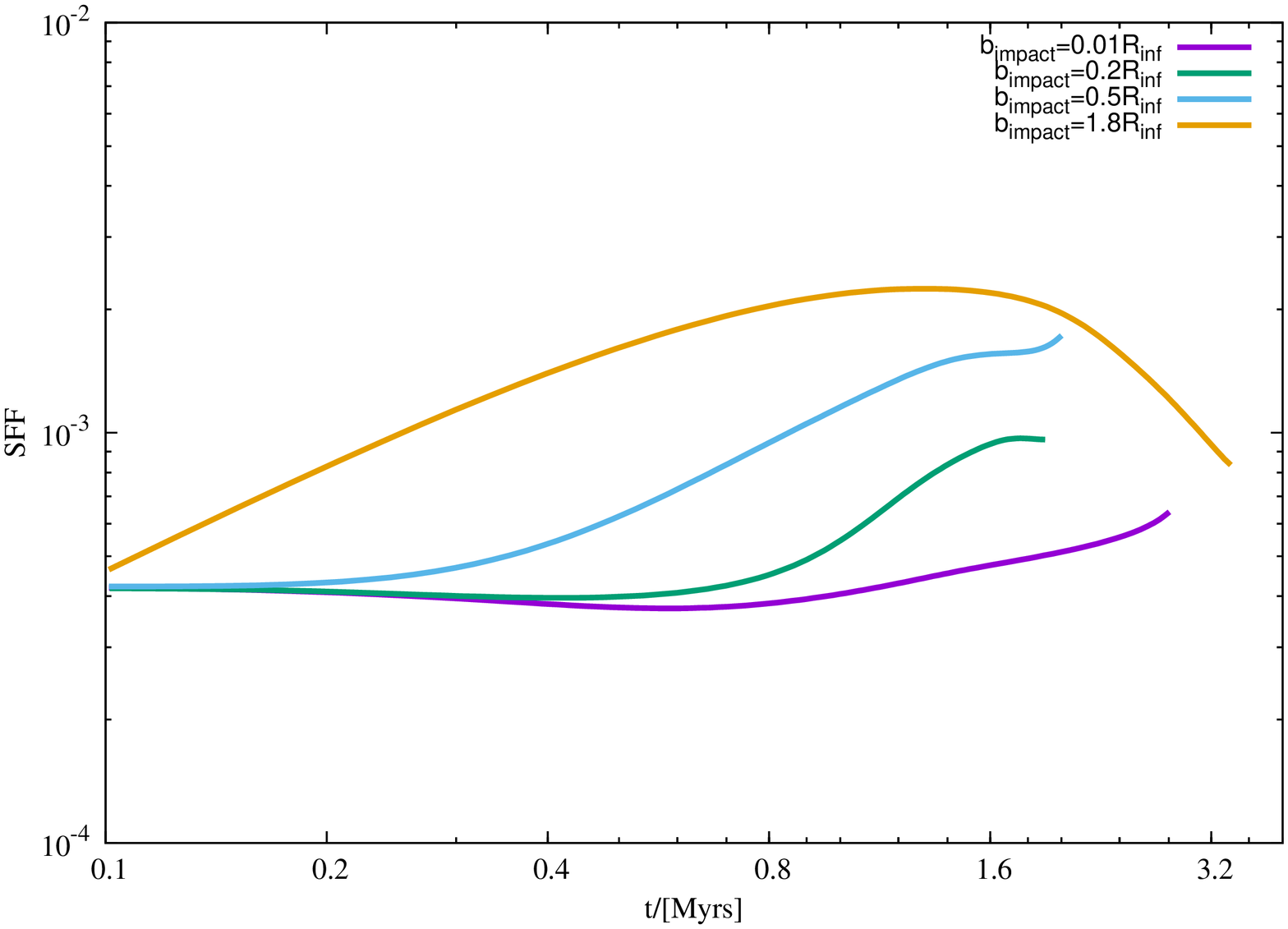}
\caption{Temporal variation of respectively the dense gas-fraction and the star-forming fraction for each choice of the impact parameter, b$_{impact}$. Bracketed quantities in the key to the figure on the upper-panel represent respectively the density threshold in units of cm$^{-3}$, and the impact parameter, b$_{impact}$, in units of the in-flow radius, $R_{inf}$. }
\end{figure}
%--------------------------------------------------------------
\textbf{Distribution of gas mass as a function of extinction}\\
Another useful diagnostic of the potential star-forming capability of a {\small MC} is the cumulative  distribution of gas mass as a function of infrared extinction, $A_{K}$. Such a plot gives us an idea about the total mass of gas available at large extinction, and therefore likely to form stars. Lada \emph{et. al.} (2009) for instance produced a similar plot for the California {\small MC}. Their plot demonstrated that the fraction of gas at large extinction fell rapidly and which could possibly help us reconcile the lack of star-formation activity in this cloud. Here we first describe the calculation of the infrared extinction, $A_{K}^{i}$, for each particle for the typical K-band wavelength of 2.2 $\mu$m. 
The mean extinction at the desired wavelength is
\begin{equation}
  \small\left\langle\exp(-A_{K}^{i})\small\right\rangle = \frac{1}{4\pi}\int \exp(-A_{K}^{i})d\Omega
\end{equation}
Keto \& Caselli (2008), $d\Omega$ being the element of solid angle subtended at the position of particle $i$. Here, $A_{K}^{i} = \Sigma_{i}Q_{K}^{i}(\nu)$, where $\Sigma_{i}$ is the average column density for the particle $i$ from the cloud surface, calculated by summing over contributions along twelve directions. The frequency dependent absorption coefficient, $Q_{K}(\nu)$, was adopted from Zucconi \emph{et al.} (2001). \\ \\
The column density here was calculated in the same manner, in the face of the slab, as for the {\small N-PDF}s shown above. Although the magnitude of the column density, and therefore the extinction, $A_{K}^{i}$, for a particle will vary if the calculation were to be done in the plane of the collision, the extant interest here is to investigate the distribution of gas across extinction-bins, and especially the variation in the fraction that ends up in the high extinction-bins. It is not our objective to deduce an exact magnitude of this fraction and so the inference drawn here is unlikely to be sensitive to any particular choice of the preferred orientation of the post-collision slab. Note that the column density used in this calculation was in the unit g cm$^{-2}$ and likewise the absorption efficiency, $Q_{K}$, originally in the units of cm$^{2}$ H$_{2}^{-1}$, was converted in to the units cm$^{2}$ g$^{-1}$ by dividing it with $\mu m_{p}$; $m_{p}$, being the mass of a proton. Now, as can be seen from the panels in Fig. 7, irrespective of the impact parameter of the collision, mass of gas at large extinction i.e., $A_{K} > 1$, falls-off relatively rapidly. \\ \\
Furthermore, these plots re-emphasise the fact that with increasing impact parameter the total mass of gas cycled into the dense pockets diminishes significantly. Also, note that as the extinction increases by an order of magnitude in the range $A_{K}\in (1,10)$, the total mass of gas also reduces gradually over intermediate values of the impact parameter, $b_{impact}$. This decline is more drastic in the fourth realisation where the inflows merely grazed past each other. By contrast, however, the nearly head-on collision between inflows appears to be the most efficient in cycling gas into the dense-phase or in other words, to large extinctions, as the decline in mass is only gradual; see plot on the top left-hand panel of Fig. 7. \\ \\
%---------------------------------------------------------
%================================================================
\begin{figure*}
\label{Figure 7}  
\vspace{1pc}
\centering
\includegraphics[angle=270,width=0.45\textwidth]{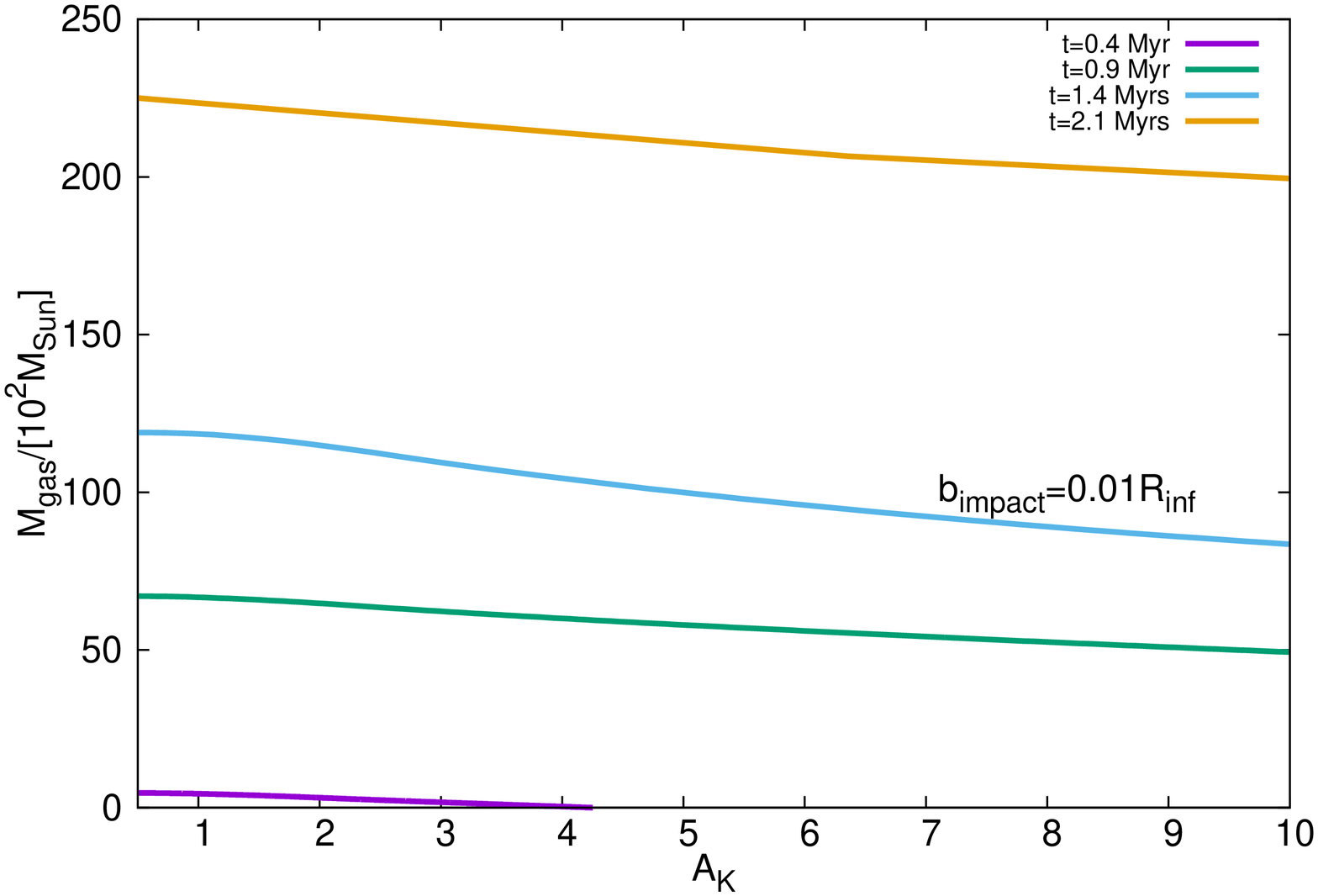}
\includegraphics[angle=270,width=0.45\textwidth]{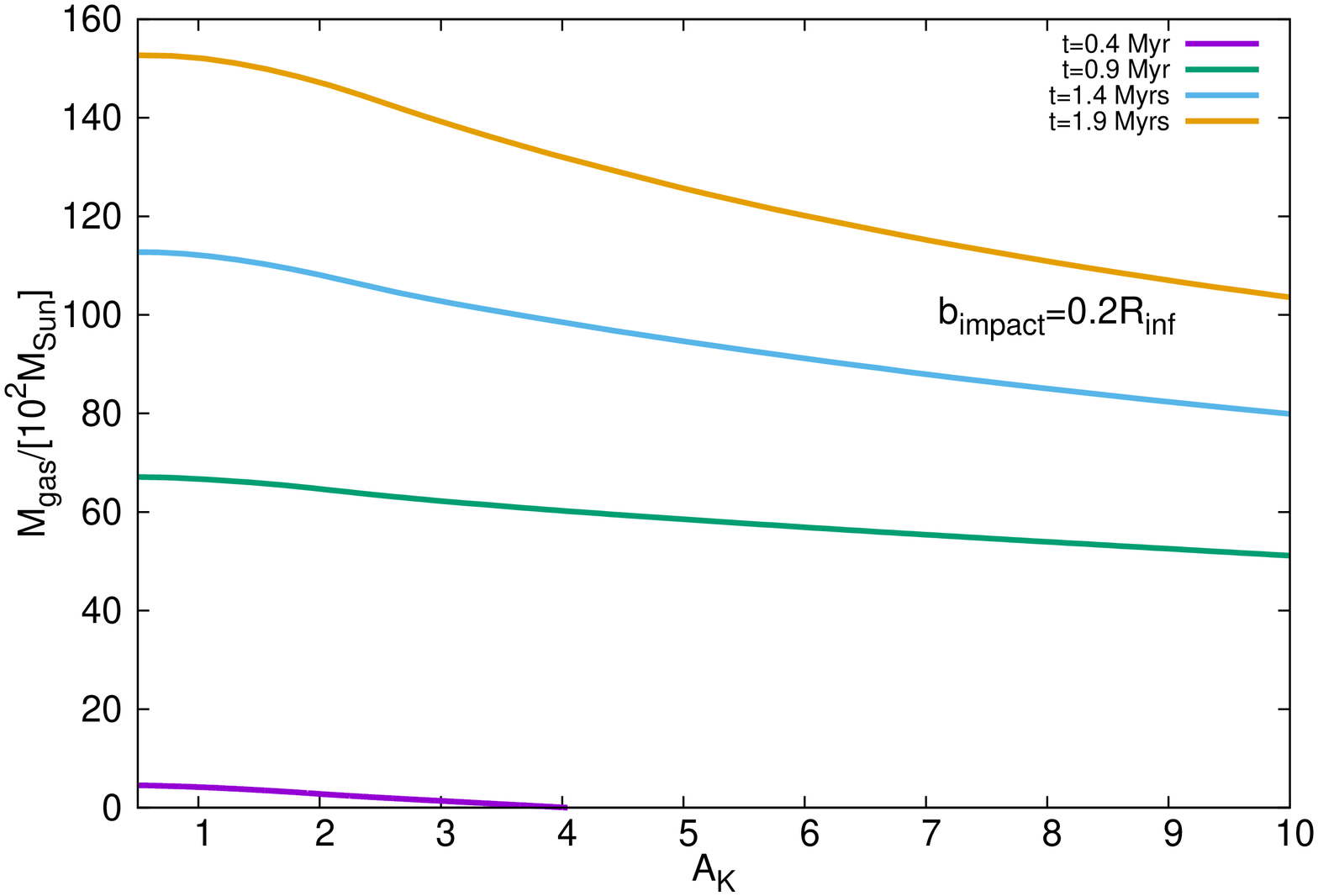}
\includegraphics[angle=270,width=0.45\textwidth]{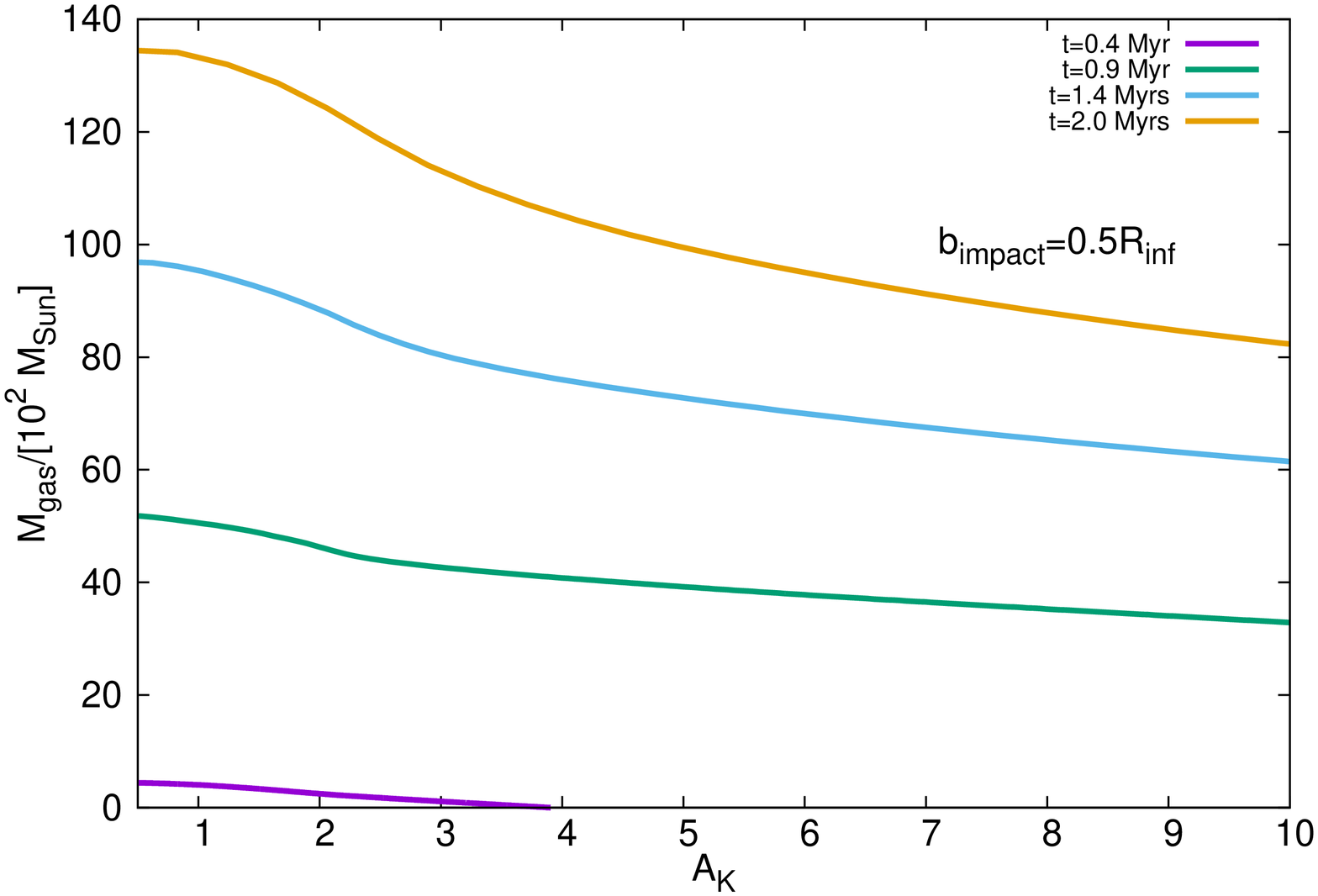}
\includegraphics[angle=270,width=0.45\textwidth]{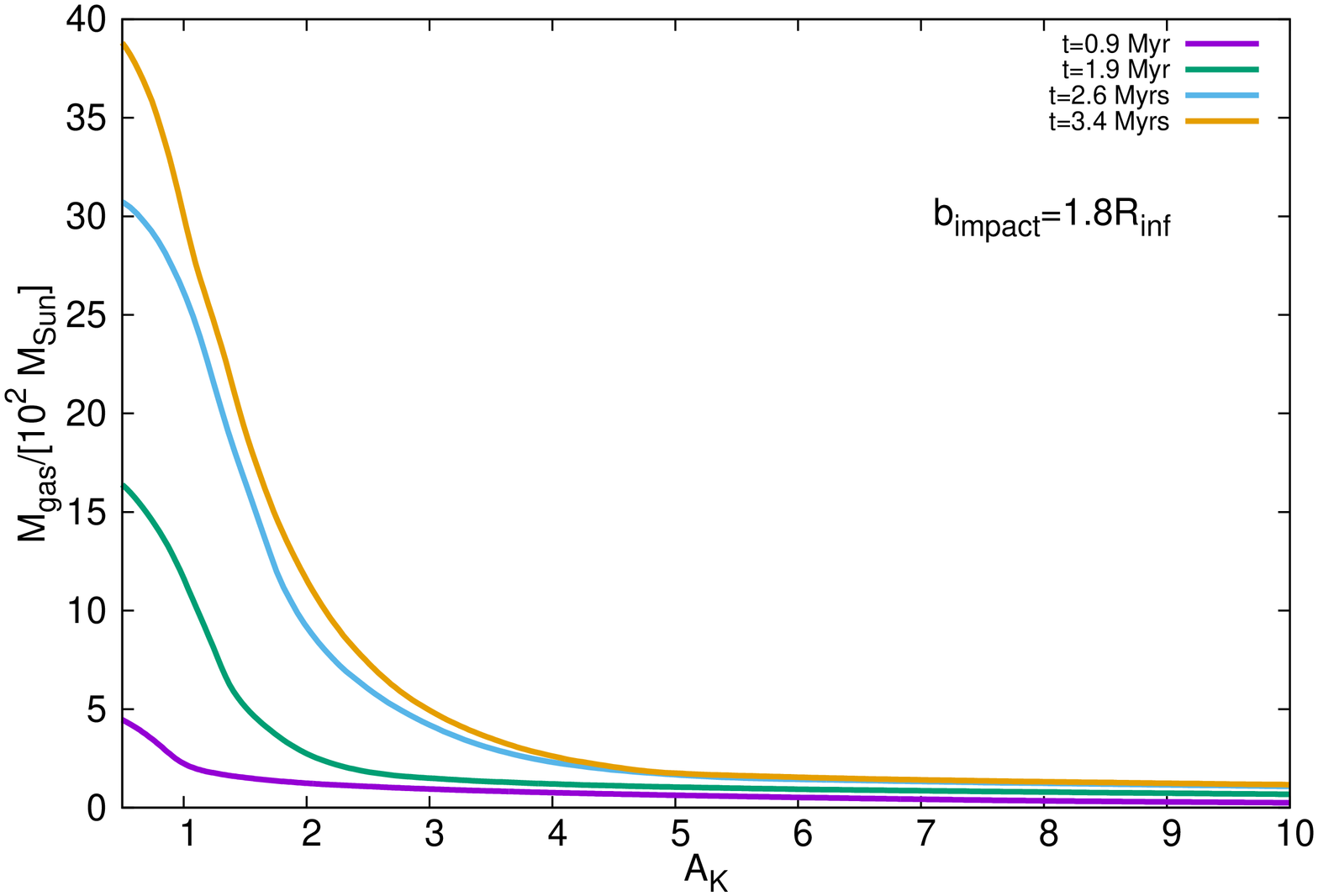}
\caption{Cumulative distribution of gas mass as a function of the extinction for the $K$-band. Barring the first realisation, the gas-mass at higher extinction ($A_{K} > 1$), diminishes falls-off relatively quickly with increasing magnitude of the impact-parameter, $b_{impact}$; this decline is the steepest for the largest magnitude of $b_{impact}$.}
\end{figure*}
%=========================================================================
\begin{figure*}
\label{Figure 8}  
\vspace{1pc}
\centering
\includegraphics[angle=270,width=0.45\textwidth]{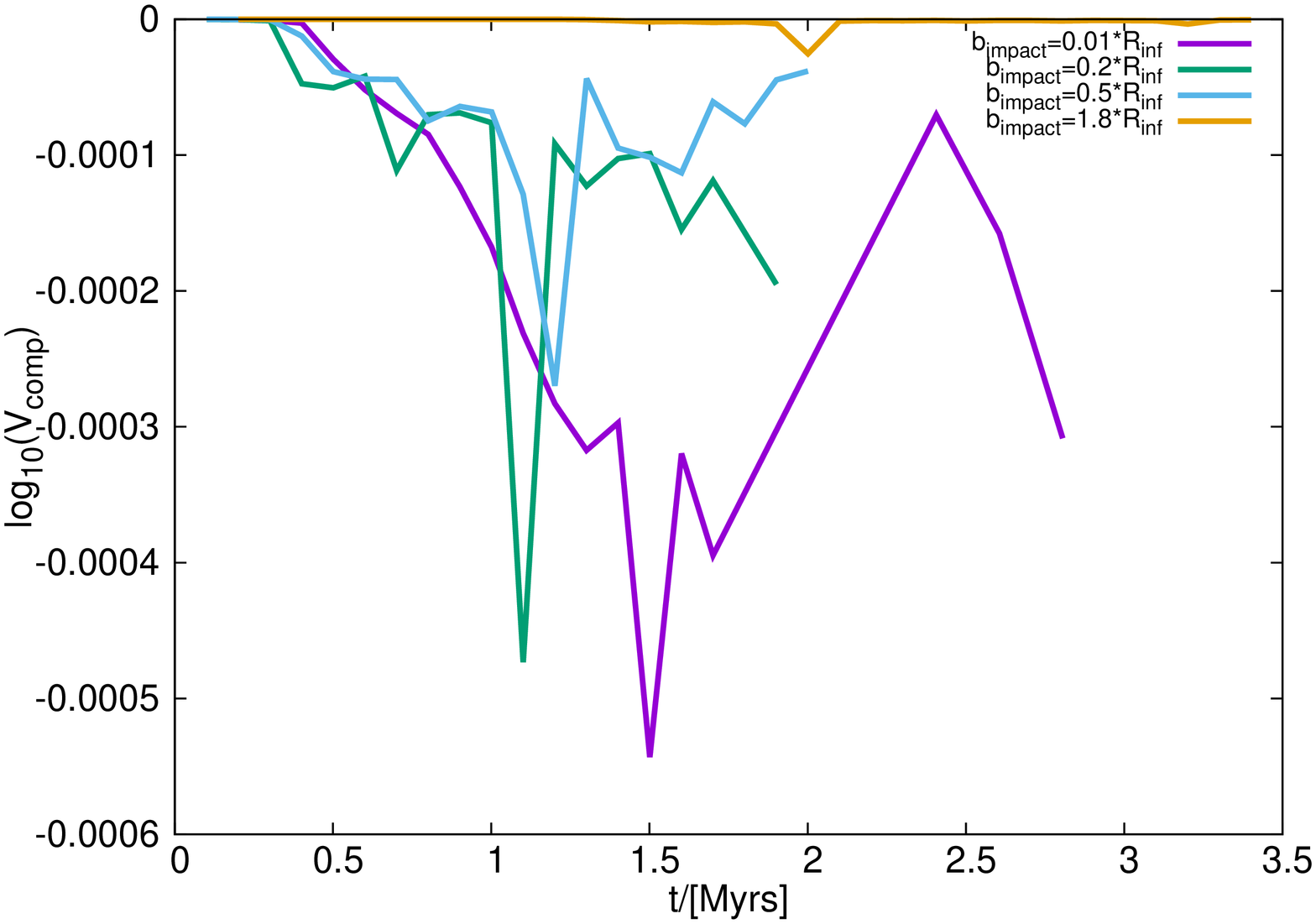}
\includegraphics[angle=270,width=0.45\textwidth]{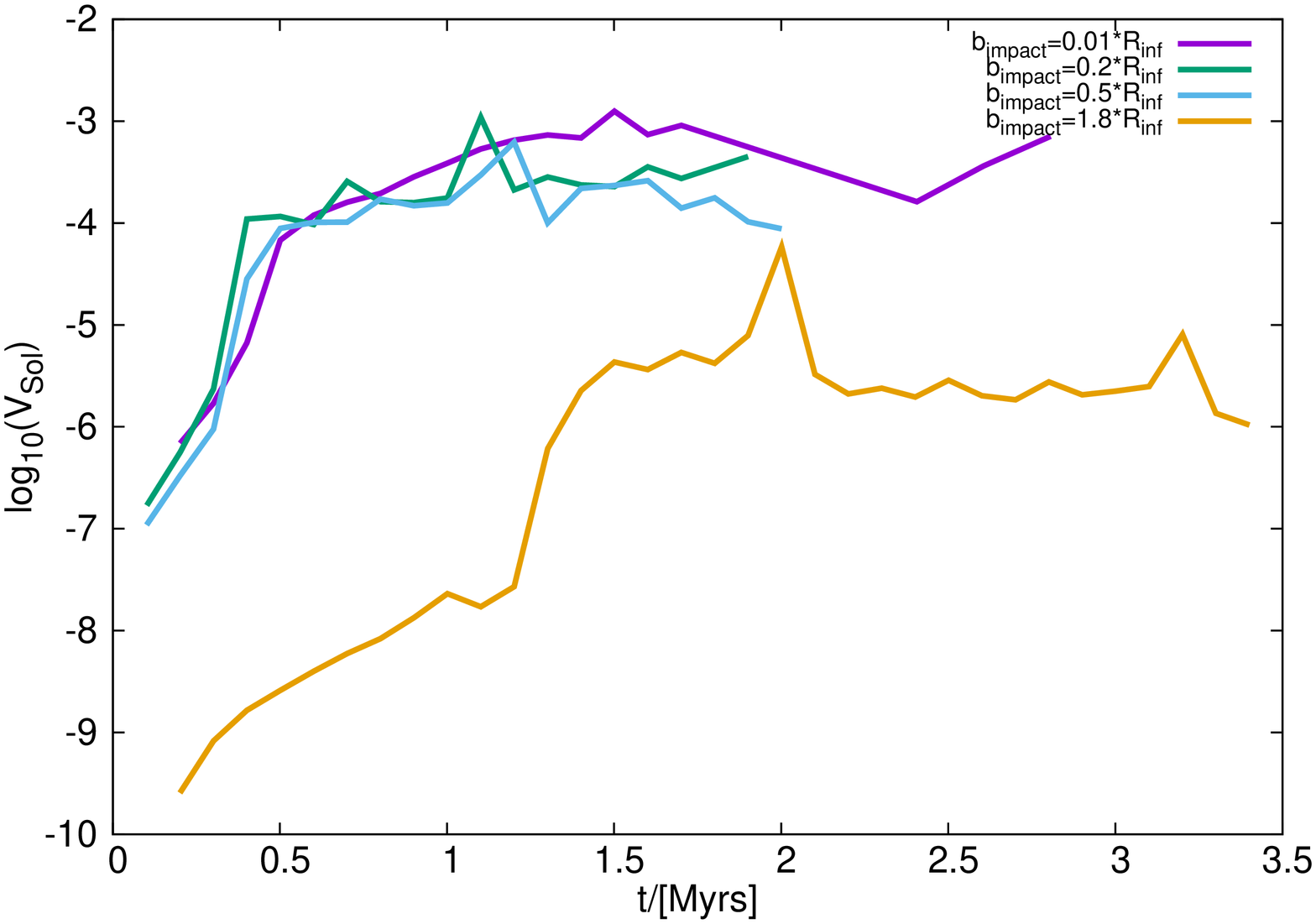}
\caption{Temporal variation of the fraction of energy in the Compressional, V$_{comp}$ $\equiv\frac{E_{comp}}{E_{tot}}$, and the Solenoidal, V$_{sol}$ $\equiv\frac{E_{sol}}{E_{tot}}$, modes of the velocity-field induced in the post-collision slab for the respective realisations is shown on the left and right-hand panel here.}
\end{figure*}
\textbf{Compressional and Solenoidal modes of the velocity field}\\
The motivation behind investigating the relative strength of the compressional and the solenoidal modes in a {\small MC} comes from the apparent dependence of the {\small N-PDF} on the nature of turbulent-forcing (e.g. Federrath \emph{et al.} 2008, Federrath \emph{et al.} 2010). These authors demonstrated that purely compressional-forcing generates an {\small N-PDF} that has a significantly larger standard deviation or in other words, purely compressional-forcing cycles gas more efficiently into the dense-phase in comparison with the solenoidal or mixed-forcing. However, we have seen in Figs. 4 and 5 above that the nature of the {\small N-PDF} is also sensitive to the impact parameter of the collision and indeed, to the magnitude of interstellar pressure, as was also argued in paper I. These inferences beg the question as to the factors likely to affect the {\small N-PDF} and furthermore, if whether the impact parameter of colliding inflows does in fact influence the strength of the compressional and solenoidal modes of the velocity field induced into the post-collision gas. \\ \\
From the Helmholtz theorem it follows that the velocity field, $\textbf{v}$, can be decomposed into two components, one of which is irrotational and the other solenoidal. Thus
\begin{equation}
  \textbf{v} = \textbf{v}_{\parallel} + \textbf{v}_{\perp},
\end{equation}
where $\textbf{v}_{\parallel}\equiv -\boldsymbol{\nabla}\phi$, and $\textbf{v}_{\perp}\equiv -\boldsymbol{\nabla}\times\textbf{A}$;
\begin{equation}
\phi(\textbf{r}_{i})\equiv\frac{1}{4\pi}\int\frac{\boldsymbol{\nabla}\cdot\textbf{v}(\textbf{r}')}{\vert\textbf{r}_{i}-\textbf{r}'\vert}d^{3}r',
\end{equation}
%----------------------------------------------
\begin{equation}
  \textbf{A}(\textbf{r}_{i})\equiv\frac{1}{4\pi}\int\frac{\boldsymbol{\nabla}\times\textbf{v}(\textbf{r}')}{\vert\textbf{r}_{i}-\textbf{r}'\vert}d^{3}r'
\end{equation}
(e.g. Arfken \& Weber 2001). Here $\textbf{r}_{i}$ is the position vector of the particle $i$ and the respective integrals in Eqs. 3 and 4 were evaluated by summing over the entire velocity field in the post-collision slab. For this purpose the slab was mapped onto a 128$^{3}$ spatial grid and particles in it were distributed on to the grid cells using an Octal spatial tree. The curl and the gradient of the underlying velocity-field were calculated on this grid and finally, the respective integrals in Eqns. 3 and 4 were evaluated by summing over contributions from individual cells after smoothing them with the standard Cubic-spline. \\ \\
Now, shown on the left and right-hand panel of Fig. 8 is the temporal variation of the fraction of energy in respectively the solenoidal component and the compressional component of the velocity field in the post-collision slab for each choice of the impact parameter. These plots demonstrate that the velocity field in the shocked-slab is by and large compressional irrespective of the impact parameter of collision and the strength of dynamic instabilities. Taken together, the observed variation of the dense gas-fraction, the {\small SFF} and the strength of compressional modes lead us to the conclusion that the pre-dominance of the compressional modes alone is not sufficient to cycle gas in to the dense-phase, though it is perhaps necessary. We note that this calculated fraction of energy is not the same as the turbulent driving parameter, $b$, often encountered in literature.  The parameter, $b$, is actually the constant of proportionality between the standard deviation of the {\small N-PDF} and the Mach number of the underlying velocity-field, the magnitude of which has been numerically estimated for different modes of turbulent forcing (e.g. Federrath \emph {et al.} 2008). By definition therefore, this parameter is not a direct proxy for the relative strength of the compressional and the solenoidal component of a turbulent velocity field.
\section{Discussion \& Conclusions}
It is now well-known that star-formation within {\small MC}s is associated with the densest pockets of gas within these clouds (e.g. Lada 1992). Consequently, one of the fundamental questions in reconciling the star-forming ability of {\small MC}s is the understanding of physical processes that cycle the relatively diffuse gas in {\small MC}s into the dense, putative star-forming pockets. Following the observational inferences from recent work by e.g. Hughes \emph{et al.} (2013), also see the review by Heyer \& Dame (2015), we demonstrated in paper I that various physical properties of {\small MC}s, including the fraction of dense-gas and the so-called star-forming fraction ({\small SFF}) in a {\small MC}, varies as the cloud evolves with time. Furthermore, we found that the extent of variation of these quantities depends on the magnitude of interstellar pressure which is also consistent with inferences drawn by the authors listed above.\\ \\
Indeed, observations of nearby {\small MC}s show that even these clouds show a significant variation in their star-forming ability as reflected by their respective star-formation efficiency. Lada \emph{et al.} (2010) demonstrated that the star-formation rate for a sample of {\small MC}s in the Solar neighbourhood shows up to an order of magnitude of variation. This is despite the fact that these clouds have comparable mass, density and the magnitude of turbulent velocity field. In this work we attempted to reconcile these observed differences with simple self-gravitating, hydrodynamic simulations of colliding flows having initially uniform density. These realisations were repeated for different values of the pre-collision impact parameter to study the possible effect on the star-forming ability of gas in the assembled cloud. These simulations were developed with a magnitude of external pressure consistent with that of the interstellar pressure found in the local neighbourhood. Now, the plot in Fig. 3 showing the temporal variation of velocity dispersion, $\sigma_{gas}$, in the post-collision slab suggests that the mere variation of the impact parameter of collision affects the magnitude of $\sigma_{gas}$. This is by virtue of the fact that for a relatively small impact parameter the {\small NTSI} appears to dominate the post-collision slab, and the turbulence injected in the slab-layers by the in-flowing gas begins to  decline only after the {\small NTSI} approaches saturation. \\ \\ 
On the other hand when the flows merely graze past each other, there is only shearing interaction between fluid-layers and the magnitude of $\sigma_{gas}$ in this case is significantly smaller. Of greater interest towards assessing the star-forming ability of a {\small MC} is its {\small N-PDF}. Following the seminal work by V{\' a}zquez-Semadeni \emph{et al.} (1995) and Padoan \emph{et al.} (1997), a log-normal  density distribution has been come to be identified with turbulence-dominated gas. Strongly self-gravitating gas, and therefore of interest from the perspective of forming stars, on the other hand develops a power-law that extends to higher densities (see for eg. Kainulainen \emph{et al.} 2009; Schneider \emph{et al.} 2012). While this is true for many clouds in the Solar neighbourhood (e.g. Lombardi \emph{et al.} 2015), it may not be the case in extreme environments such as that encountered in the Galactic {\small CMZ} (e.g. Rathborne \emph{et al.} 2014). \\ \\
For instance, the {\small N-PDF} for the \emph{Brick} is log-normal despite having a relatively large average density, typically on the order of $\sim 10^{4}$ cm$^{-3}$. Even in the nearby {\small MC}s the {\small N-PDF}s show a considerable variety; Pokhrel \emph{et al.} (2016) for instance, reported that the power-law tail for some regions in the Mon R2 {\small MC} exhibits a break. In such clouds the tail at the high-density end can in fact be fitted with two power-laws. Similarly, Stutz \& Kainulainen (2015) showed that the power-law slope at the high-density end of the {\small N-PDF} in the Orion varied spatially, i.e., regions showing evidence of star-formation exhibited a relatively shallow slope. Also, Hill \emph{et al.} (2011) reported more complex {\small N-PDF}s for the Rosette {\small MC}. In paper I we demonstrated the variation of the {\small N-PDF} as a function of the magnitude of the external-pressure. Plots in Figs. 4 and 5 show that even for a magnitude of pressure typically found in the Solar neighbourhood, the mere variation of the impact parameter of colliding-flows that could precede clouds, also alters the nature of the {\small N-PDF}. In the case where the inflows collided almost head-on the {\small N-PDF} is qualitatively similar to the ones deduced for typical star-forming clouds such as the Orion {\small MC} or the Taurus (e.g. Kainulainen \emph{et al.} 2009, Lombardi \emph{et al.} 2015), or sometimes exhibits a break in the high-density tail similar to that reported by Pokhrel \emph{et al.}
 The {\small N-PDF}s shown in Fig. 5 also demonstrate that with increasing impact parameter the slope of the power-law tail at the high-density end tends to steepen. In particular, 
 the {\small N-PDF}s derived here for realisation 4 where the flows merely graze past each other appear qualitatively similar to the fits provided by Lombardi \emph{et al.} to the {\small N-PDF}s deduced by them for clouds such as the Perseus, Pipe or the California. All of these latter clouds exhibit little star-forming activity.  \\ \\
The potential star-forming ability of a cloud can also be quantitatively assessed by investigating into the fraction of gas cycled into the dense-phase. The latter realisation, realisation 4, is most inefficient in transforming gas into the dense-phase as is reflected by the relatively small dense-gas fraction in the plot on the upper-panel of Fig. 6. Similarly, the {\small SFF} shown on the lower-panel of Fig. 6, for this latter case declines rapidly to less than a percent while that for the remaining three realisations, can be seen to rise steadily. In fact, Orkisz \emph{et al.} (2017) attribute the observed poor efficiency of star-formation in the Orion B cloud to the impact of large-scale differential motion causing the gas within the cloud to shear. \\ \\
The next set of plots in Fig. 7 showing the cumulative distribution of gas mass as a function of visual-extinction (actually calculation for the K-band extinction), is also quite revealing from the perspective of the star-forming ability of a cloud. Similar plots were deduced by Lada \emph{et al.} (2010) for an ensemble of {\small MC}s in the Solar neighbourhood. Clouds such as the Pipe, Perseus, California and the Lupus 1, 3 \& 4 known for their poverty in forming stars, showed mass-distribution very similar to that for clouds assembled via collisions having a large impact parameter (realisation 4 here), where there is very little mass at large extinction. By contrast, the mass-distribution for the cloud assembled in realisation 1 where the collision was almost head-on is similar to that for {\small MC}s such as the Orion or Ophiuchus that form stars more efficiently. It is of course true that stellar feedback in these latter clouds must also contribute to star-formation in these latter clouds. \emph{We therefore restrict ourselves to the inference that a pristine cloud assembled via head-on or near head-on collision is likely to cycle a larger fraction of gas to higher extinctions ($A_{K} > 1$), which could then possibly become future birth-sites for stars.}  It is indeed interesting that a relatively simple self-gravitating, hydrodynamic model such as the one discussed here can possibly reconcile the observed variations in the distribution of gas mass as a function of visual extinction in nearby {\small MC}s. \\ \\
Finally, we examined the possible correlation between the fraction of gas cycled into the dense-phase and the relative strength of compressional/solenoidal modes in the velocity-field induced in the assembled cloud. The propensity of compressional modes towards assembling putative star-forming gas was demonstrated by Federrath \emph{et al.} (2008). They showed numerically that gas dominated by compressional turbulence, as indicated by the strength of the so-called driving-parameter, $b$, tended to produce {\small N-PDF}s with a relatively large standard-deviation, and possibly with a power-law extension at higher densities. By contrast, dominance of solenoidal modes tended to produce log-normal  {\small N-PDF}s. In other words, compressional modes were found to be more propitious towards star-formation. This conclusion was reinforced by Federrath \emph{et al.} (2010), who further demonstrated that prestellar cores formed on the so-called sonic-scale, i.e., the length-scale where turbulence made the transition from being supersonic to sub-sonic. Furthermore, Federrath (2013) proposed a new star-formation law according to which the observed variations in the star-formation rate across {\small MC}s can be reconciled by accounting for the possible variation in the Mach number of the turbulent velocity field prevailing in those clouds. \\ \\
A similar suggestion was also made by Anathpindika (2013), who following a semi-analytic calculation demonstrated that along with the Mach number of the turbulent velocity-field, the length-scale on which turbulent energy is injected in a cloud could affect the distribution of core masses. Kainulainen \emph{et al.} (2014) argued that the star-formation efficiency of many nearby {\small MC}s can be reconciled if the turbulent velocity-field threading these clouds is relatively non-compressive. By contrast, the usually filament-like infrared dark clouds ({\small IRDC}s), must be dominated by strongly compressive turbulent velocity-fields. The Orion B is another example of a molecular cloud that shows a relatively poor {\small SFE} (e.g. Lada \emph{et al.} 2010). In fact, Megeath \emph{et al.} (2016) argue that it has the lowest {\small SFE} among all the clouds in a radius of $\sim$500 pc. Orkisz \emph{et al.} (2017) attribute this poverty in forming stars to the dominance of solenoidal modes on the large-scale which is consistent with the hypothesis that solenoidal modes in general inhibit star-formation. Similarly, Federrath \emph{et al.} (2016) also argued that the observed poor star-formation in the \emph{Brick} could be due to the dominance of solenoidal modes in that cloud. Federrath (2015) also demonstrated that the widely reported filament-width of $\sim$0.1 pc could be reconciled with a mixture of self-gravity and turbulence. This typical filament-width, according to the author is on the order of the sonic length-scale. The author also argued that the filament width tended to be narrower in purely self-gravitating simulations without turbulence. \\ \\
Now, despite the apparent success of the variously reported turbulence-based models with/out the magnetic field, the matter as to the origin of turbulence and its strength in {\small MC}s is an open question. Furthermore, the apparent efficiency of solenoidal modes in suppressing star-formation (e.g. Federrath \emph{et al.} 2008, 2010), as reflected by a relatively poor star-formation efficiency, tends to emphasise the importance of compressional modes over the former. Though from these works it is unclear if these modes also carried bulk of the energy in the turbulent velocity-field. Plots shown in Fig. 8, however, suggest a stronger compressional component need not imply a greater star-formation efficiency. \emph{Evidently, the strong dominance of compressional modes may not always ensure that a proportionally large fraction of gas will be cycled to the typical star-forming densities.} Also, it is unclear how a pristine cloud might come about to be dominated by solenoidal modes, for we have seen above that the assembled cloud is largely dominated by compressional modes even when it is strongly sheared as in the case where 
the impact parameter of inflows assembling it is so large that the flows merely graze past each other. It is therefore evident that predominance of compressional modes is insufficient for enhancing the dense-gas fraction, or indeed the {\small SFF} in a cloud. It is possible that solenoidal modes are injected by expanding shells triggered by ongoing star-formation in a cloud. Furthermore, since the turbulence-modulated star-formation model discussed in the preceding paragraphs assumes the Mach number as an \emph{ansatz}, it cannot shed much light on the impact of external pressure i.e. the magnitude of interstellar pressure on the star-forming ability of a cloud. \\ \\
Clouds in the Solar-neighbourhood present an interesting case in point. \emph{The physical conditions prevailing in these clouds are mutually similar if not identical (e.g. Lada \emph{et al.} 2009), and so they are unlikely to exhibit variations in the average Mach number as large as those warranted by the above discussed turbulence-based model in order to reconcile the observed variation in their respective star-formation efficiency.} While one could invoke the magnetic field to explain the sluggish nature of star-formation in some of these clouds, the strength of magnetic field necessary to inhibit star-formation would be significantly higher than that typically found in {\small MC}s (e.g. Kirk \emph{et al.} 2006). Crucially, there is also dearth of observational evidence to corroborate the existence of strongly magnetised clouds.
\subsection{Limitations of this work}
The work presented in this article suffers from three shortcomings. First, we have not included the magnetic field in these realisations. The magnetic field is likely to stabilise the post-collision slab against dynamical instabilities and thus affect the fraction of gas cycled into the dense-phase. Furthermore, it would also be interesting to investigate if the orientation of the magnetic field in the pre-collision flows relative to the plane of collision has any bearing on the fraction of gas cycled into the dense-phase and indeed, on the nature of the {\small N-PDF}. Despite this shortcoming, our principal conclusion that relative poverty of putative star-forming gas in some {\small MC}s in the Solar neighbourhood can possibly  be reconciled within the paradigm of cloud-formation via collision between gas-flows, is fairly robust. Second, realisations discussed in this work could not be followed up to typical prestellar densities. Consequently, we are unable to address directly the question about core-formation efficiency ({\small CFE}) in a pristine cloud, and its dependency on ambient environment of the natal cloud. Instead, we have used the dense gas-fraction and the star-forming fraction as proxies to asses the potential of the assembled cloud to form prestellar cores. We will address this problem more directly in our forthcoming article. \\ \\
Finally, we also appreciate the limitations associated with this set-up. Although we have used the findings of Kuno \& Nakai (1997) to model the pre-collision gas-flows as cylinders of finite length, the lack of any detection of galactic shocks in the {\small M51} by these authors may be cited as a possible criticism of our model where cloud-formation is explicitly induced via a collision between the initially cylindrical flows. The non-detection of any galactic-shock by Kuno \& Nakai is interesting because although the gas-streamlines are consistent with the density wave theory, the shocks predicted by this theory could not be detected. This non-detection, the authors suggested, was possibly due to the particle-like behaviour of gas-packets, i.e., streamlines of molecular gas could be rather composed of discrete packets of gas. Alternatively, these authors also suggested that self-gravity of the {\small M51} arms could be influencing the gas-dynamics. \\ \\
This is a possibility that we have not explored here. Although our simulations are self-gravitating, we have not imposed an external potential and so, we are unable to discuss the impact of the spiral-potential of the galactic-arm on cloud-formation. It is quite likely that the external potential might partially mitigate the impact of the post-collision shock. If indeed this is the case then with the set-up used here, we could be seeing an exacerbation of the impact of the shock-induced shear between the layers of gas in the post-collision slab. 
It is also likely that the effect of the shell-instability is probably magnified in the set-up being used here. Perhaps the dramatic/severe effects of the shell-instability observed in the simulations reported here may not in reality be as severe for a gas-body of arbitrary geometry. However, we have previously experimented with spherical geometry and the found that the post-collision interface is planar, slab-like and unstable to the shell-instability \textbf{(Anathpindika 2009, Anathpindika 2010)}. Thus, it appears that a set-up invoking some sort of convergence of gas-flows will likely suffer from this difficulty, though perhaps the post-collision interface/sheet might be more stable against the shell instability in the presence of an external potential.\\ \\
Furthermore, the set-up used in this work is relatively simple. It does not include the possible impact of the fractal nature of in-flows and the large-scale galactic shear. Carroll-Nellenbeck \emph{et al.} (2014), for instance compared the evolution of the post-collision cloud forming out of initially uniform gas-flows and flows that were initially fractal. They found, the post-collision cloud in the former case fragmented rapidly due to the early onset of the hydrodynamic instability and produced a lower core-formation rate with predominantly low-mass cores in comparison with the latter. It is difficult to include the impact of galactic shear in a simplified model such as this one. However, the colliding-flow model has been frequently used in the past and has provided a significant insight into the formation of {\small MC}s and the variation of their physical properties. In this work we have explored questions not attended so far.
\section{Summary}
The star-formation efficiency of a {\small MC} depends on the fraction of gas cycled into the dense-phase in a cloud. In a pristine cloud, this fraction is likely determined by factors at play during the assembly of a cloud. It is known that the magnitude of external pressure acting on a {\small MC} does affect the star-forming fraction ({\small SFF}) in a cloud. However, observations of {\small MC}s in the Solar neighbourhood show that the {\small SFF} can vary in clouds despite the similarity in their physical characteristics, and in the ambient conditions. In this  work we attempted to reconcile this observed variation. For a choice of external pressure consistent with the estimated magnitude of interstellar pressure in the Solar neighbourhood, we varied the impact parameter of the inflows that assemble a cloud. \\ \\
We found that although a relatively large impact parameter stabilises the post-collision slab against the thin-shell instability, it also restricts the ability of cycling the gas into the dense phase. Consequently, the dense gas fraction and the {\small SFF} tends to decline with increasing impact parameter. Furthermore, the slope of the high-density tail of the {\small N-PDF} tends to steepen significantly with increasing impact parameter.
 This suggests, strongly sheared clouds are likely to form stars less efficiently than others despite having similar physical properties such as gas-mass, density and velocity dispersion. Contrasting star-forming histories exhibited by mutually similar {\small MC}s such as those reported by Lada \emph{et al.} (2009) could therefore be attributed to the effect of shear. \emph{We also note that for shear to stabilise a cloud against self-gravity, the cloud need not necessarily be located in the inter-arm region of a galaxy since off-centre collision between the flows responsible for assembling a cloud in a galactic-arm could also induce shear. This latter possibility is plausible in a turbulent interstellar medium}.\\ \\
The simple self-gravitating, hydrodynamic realisations discussed in this work offer a relatively straightforward explanation to the observed variation of the {\small SFE} in mutually similar {\small MC}s. We have demonstrated that the mere dominance of the compressional modes in a pristine cloud, even if that cloud is strongly sheared, is not sufficient to ensure a large {\small SFF}. This conclusion is at odds with the turbulence-regulated star-formation model. According to this latter model a turbulent velocity-field largely dominated by solenoidal modes could possibly reconcile the observed variation in {\small SFE}s of {\small MC}s. However, it is not clear how solenoidal modes could be injected in a virgin cloud. Galactic shear could possibly inject solenoidal modes in a cloud which would then suggest that clouds located only in the inter-arm region should exhibit a  poor efficiency of converting gas into the dense-phase. However, we know that several clouds even in the local neighbourhood have a relatively small fraction of putative star-forming gas. We have shown in this work that a cloud assembled by interacting gas-flows are largely dominated by compressional modes. Our model also circumscribes the discomforting question with regard to the strength of the magnetic field necessary to render star-formation less efficient in some clouds as compared with others.

\section*{Acknowledgements}
{\small SVA} is supported by the grant {\small YSS/2014/000304} of the Department of Science \& Technology, Government Of India, under the Young Scientist Scheme. This sequel is a result of the collaboration initiated with financial assistance in the form of a Royal Astronomical Society grant awarded October 2015. The author gratefully acknowledges useful discussions with Charlie Lada and generous financial support in the form of a Visiting fellowship of the Institute for Computation at the Harvard-Smithsonian Centre for Astrophysics. We gratefully thank the bwGRiD project \footnote{BwGRiD (http://www.bw-grid.de), member of the German D-Grid initiative, funded by the Ministry for Education and Research (Bundesministerium fuer
Bildung und Forschung) and the Ministry for Science, Research and Arts Baden-Wuerttemberg (Ministerium fuer Wissenschaft, Forschung und Kunst Baden-Wuerttemberg)} for the computational resources that were used to develop simulations discussed in this work. {\small RK} acknowledges financial support via the Emmy Noether Research Group on “Accretion Flows and Feedback in Realistic Models of Massive Star Formation” funded by the German Research Foundation ({\small DFG}) under grant no. {\small KU 2849/3-1}.

\bsp

\label{lastpage}

\end{document}